\documentclass[pre,preprint,longbibliography]{revtex4-1}%
\usepackage{amssymb}
\usepackage{amsfonts}
\usepackage{amsmath}
\usepackage{xcolor}
\usepackage{ulem}
\usepackage{graphicx}%
\providecommand{\U}[1]{\protect\rule{.1in}{.1in}}
\begin{document}
\title{Classical Density Functional Theory in the Canonical Ensemble}
\author{James F. Lutsko}
\homepage{http://www.lutsko.com}
\email{jlutsko@ulb.ac.be}
\affiliation{Center for Nonlinear Phenomena and Complex Systems CP 231, Universit\'{e}
Libre de Bruxelles, Blvd. du Triomphe, 1050 Brussels, Belgium}

\begin{abstract}
Classical density functional theory for finite temperatures is usually
formulated in the grand-canonical ensemble where arbitrary variations of the
local density are possible. However, in many cases the systems of interest are
closed with respect to mass, e.g. canonical systems with fixed temperature and
particle number. Although the tools of standard, grand-canonical density
functional theory are often used in an ad hoc manner to study closed systems,
their formulation directly in the canonical ensemble has so far not been
known. In this work, the fundamental theorems underlying classical DFT are
revisited and carefully compared in the two ensembles showing that there are
only trivial formal differences. The practicality of DFT in the canonical
ensemble is then illustrated by deriving the exact Helmholtz functional for
several systems: the ideal gas, certain restricted geometries in arbitrary
numbers of dimensions and finally a system of two hard-spheres in one
dimension (hard rods) in a small cavity. Some remarkable similarities between
the ensembles are apparent even for small systems with the latter showing
strong echoes of the famous exact of result of Percus in the grand-canonical ensemble.

\end{abstract}
\date{\today }
\maketitle

\section{Introduction}

Density functional theory (DFT) is a powerful reformulation of equilibrium
statistical mechanics that has found applications throughout physics. The most
well-known version of DFT is that for quantum systems at zero temperature
(qDFT) which is a fundamental tool used in applications in materials science,
chemistry and physics\cite{qDFT,ABC}. Conceptually related, but quite
different in practice, is classical DFT (cDFT) for systems at non-zero
temperature (see, e.g. \cite{Evans1979, lutsko:acp}). Recently, quantum DFT at
non-zero temperatures has drawn increasing attention as well (see, e.g.
\cite{WDM,Smith2018}). All three varieties have the same conceptual
structure:\ one proves that there is one-to-one mapping between external
applied fields and the local number density. A corollary of this proof is the
existence of a functional of the one-body density which is minimized by the
equilibrium density distribution. At zero temperature the value of the
functional evaluated at its minimum is the ground-state energy of the system
whereas for the finite temperature cases it is the grand-canonical free
energy. In general, this energy functional is not known and applications
depend on carefully constructed approximate functionals which are usually
constrained by various exact limits and scaling relations, in the quantum
case, or by certain specific exact results in the classical case.

An aspect of DFT that has always caused confusion is the fact that the
classical theorems for finite temperature systems are proven in the grand
canonical ensemble\cite{MerminDFT}. One reason for this is simply that it is
easier, at the formal level, to work in the grand ensemble than it is under
the constraint of fixed particle number that is required for the canonical
ensemble. In typical DFT applications, the distinction is often of little
practical importance since the ensembles are equivalent in the thermodynamic
limit. However, in applications on small systems, in particular, the
differences between the ensembles can be qualitatively large. It is sometimes
thought that one can simply minimize the grand-canonical energy functional
under the constraint of a fixed number of particles and thereby get the
canonical result but this is not true :\ this only fixes the average number of
particles in the grand-canonical calculation and does not eliminate the effect
of particle number fluctuations which do not exist in the true canonical
system. Besides small systems, another important motivation for wanting a
canonical version of DFT is that dynamical models often require as input a
free energy functional and it is natural to use the sophisticated functionals
developed in DFT\cite{Lutsko_HCF}. However, dynamical models are almost always
formulated for canonical systems (e.g. starting from the Liouville equation)
and so the use of grand-canonical energy functionals is always open to
question. This also holds true for Dynamical Density Functional
Theory\cite{Duran,Lowen}, although the point is often not discussed.

Over the years, there have been a number of proposals coming from the
statistical mechanics community\cite{Lebowitz}, the quantum condensed matter
community\cite{KGV} and the classical DFT community\cite{GWRE, Schmidt} for
extracting more or less exact canonical results from grand-canonical
calculations (or in general, results in one ensemble from calculations in
another). However, the direct formulation of finite-temperature DFT in the
canonical ensemble seems to have been little explored until now. A notable
exception is the work of White and Velasco\cite{W1} and of White and
Gonz\'{a}lez\cite{W2}. In these papers, the formalism is discussed without
however without giving constructive derivations of the variational principle
and without giving exact results beyond the basic example of the ideal gas. One
should also mention work by Ashcroft\cite{Ashcroft} who similarly explores some of
the formal statistical mechanics of cDFT in the canonical ensemble, but again without any applications.  
The perspective of the present work differs from previous discussions in two
ways. First, by comparing constructive proofs of the formalism in the two ensembles, it
becomes clear that there is no overwhelming difference in the formalism of DFT
in the open and closed ensembles. This is not to say that there are {\it no} differences,
as sometimes implied in formal discussions (see e.g. Parr and Wang\cite{Parr}) but,
rather, that the differences are easily accounted fo. Second, the present discussion differs in further illustrating this point
by development of nontrivial exact results in the canonical ensemble mirroring those
already known in the grand-canonical ensemble. In the next Section, the basic
theorems of Mermin and Evans that underlie finite-temperature (classical) DFT
are revisited by following them step-by step in both the canonical and the
grand canonical ensembles. The result is that there is virtually no difference
aside from the fact that in the canonical ensemble the relation between
external fields is not one-to-one, as in the grand-canonical ensemble, but
rather the local density maps uniquely onto an affine family of external
fields, which makes little practical difference. This formal similarity is
exploited in the third and fourth sections where some exact results are given.
First, the rather trivial example of the ideal gas, previously known from the
work of White et al\cite{W1,W2}, is re-derived from the present perspective.
Second, the exact functionals for various collections of small cavities in
arbitrary dimensions are determined and compared to the corresponding
grand-canonical results. Third, the highly non-trivial problem of hard-rods in
one dimension is discussed. In the grand-canonical ensemble, the exact
functional for this system was given by Percus\cite{Percus1, Percus2, Percus3}
and these have since played in a central role in the development of
cDFT\cite{lutsko:acp}. The problem is in some ways more difficult in the
canonical ensemble and here only the special case of two hard-rods in a small
cavity is worked out. Nevertheless, it is possible to construct the exact
solution and in the limit that the cavity becomes just large enough to hold
two rods, the functional is very similar to Percus' general result, which is
quite surprising given that one is in some sense making the worse comparison
possible - a grand-canonical result to a canonical result for a very small
system. The paper concludes with a discussion of the implications of these results.

\bigskip

\section{DFT in the canonical ensemble}

\subsection{Notation}

Consider a system of $N$ particles with positions and momenta $\mathbf{q}_{i}$
and $\mathbf{p}_{i}$ respectively, for $1\leq i\leq N$. The collection of all
phases will be denoted as $\Gamma^{\left(  N\right)  }$ and the Hamiltonian
for the $N$-particle system is $\widehat{H}_{N}$where the caret means that the
quantity depends on the phases $\Gamma^{\left(  N\right)  }$. The systems are
subject to an external one-body field $\phi_{\mathbf{r}}$ so that
\begin{equation}
\widehat{H}_{N}\left[  \phi\right]  =\widehat{H}_{N}\left[  0\right]
+\sum_{i=1}^{N}\phi_{\mathbf{q}_{i}}%
\end{equation}
where the square brackets denote a functional dependence and, in order to keep
separate the function and functional dependencies, positions and momenta will
be denoted as subscripts so that what is written here as $\phi_{\mathbf{q}%
_{i}}$ would normally be written as $\phi\left(  \mathbf{q}_{i}\right)  $. In
the following, I will give the equations for each step of the arguments
simultaneously for the grand-canonical (GC) and canonical (C) ensembles so
that the close similarity - and important differences are apparent.

\subsection{Definitions}

Let
\begin{align}
\widehat{f}_{N}\left[  \phi\right]   &  =\frac{1}{Z_{N}\left[  \phi\right]
N!h^{ND}}\exp\left(  -\beta\widehat{H}_{N}\left[  \phi\right]  \right)
,\;\;\text{(C)}\\
\widehat{f}_{N\mu}\left[  \phi\right]   &  =\frac{1}{\Xi_{\mu}\left[
\phi\right]  N!h^{ND}}\exp\left(  -\beta\left(  \widehat{H}_{N}\left[
\phi\right]  -\mu N\right)  \right)  ,\;\text{(GC)}\nonumber
\end{align}
where $h$ is Planck's constant, $\beta=1/k_{B}T$ is the inverse temperature
and $k_{B}$ is Boltzmann's constant, and $\mu$ is the chemical potential. For
the canonical case, this is just the usual equilibrium distribution while for
the grand-canonical case, it is the $N-$body contribution to the full
distribution. The canonical and grand canonical partition functions are%
\begin{align}
Z_{N}\left[  \phi\right]   &  =\frac{1}{N!h^{ND}}\int\exp\left(
-\beta\widehat{H}_{N}\left[  \phi\right]  \right)  d\Gamma^{\left(  N\right)
},\;\;\text{(C)}\\
\Xi_{\mu}\left[  \phi\right]   &  =\sum_{N=0}^{\infty}\frac{1}{N!h^{ND}}%
\int\exp\left(  -\beta\left(  \widehat{H}_{N}\left[  \phi\right]  -\mu
N\right)  \right)  d\Gamma^{\left(  N\right)  },\;\text{(GC)}\nonumber
\end{align}
and the corresponding free energies are
\begin{align}
\beta A_{N}\left[  \phi\right]   &  =-\ln Z_{N}\left[  \phi\right]
,\;\;\text{(C)}\\
\beta\Omega_{\mu}\left[  \phi\right]   &  =-\ln\Xi_{\mu}\left[  \phi\right]
,\;\;\text{(GC).}\nonumber
\end{align}

The central quantity in the analysis is of course the average local density.
It is defined in terms of the microscopic density,%
\begin{equation}
\widehat{\rho}_{N\mathbf{r}}=\sum_{i=1}^{N}\delta\left(  \mathbf{r}%
-\mathbf{q}_{i}\right)  ,
\end{equation}
as
\begin{align}
\rho_{N\mathbf{r}}\left[  \phi\right]   &  =\int\widehat{\rho}_{N\mathbf{r}%
}\widehat{f}_{N}\left[  \phi\right]  d\Gamma,\;\;\text{(C)}\label{explicit}\\
\rho_{\mu\mathbf{r}}\left[  \phi\right]   &  =\sum_{N=0}^{\infty}\int
\widehat{\rho}_{N\mathbf{r}}\widehat{f}_{N\mu}\left[  \phi\right]
d\Gamma,\;\;\text{(GC).}\nonumber
\end{align}
Notice that in terms of the density one has that the Hamiltonian can be
written as
\begin{equation}
\widehat{H}_{N}\left[  \phi\right]  =\widehat{H}_{N}\left[  0\right]
+\int\widehat{\rho}_{N\mathbf{r}}\phi_{\mathbf{r}}d\mathbf{r}%
\end{equation}
and as a consequence, one verifies from the definitions that
\begin{align}
\frac{\delta\beta A_{N}\left[  \phi\right]  }{\delta\beta\phi_{\mathbf{r}}}
&  =-\frac{\delta\ln Z_{N}\left[  \phi\right]  }{\delta\beta\phi_{\mathbf{r}}%
}=\rho_{N\mathbf{r}}\left[  \phi\right]  ,\;\;\text{(C)}\label{D}\\
\frac{\delta\beta\Omega_{\mu}\left[  \phi\right]  }{\delta\beta\phi
_{\mathbf{r}}}  &  =-\frac{\delta\ln\Xi_{\mu}\left[  \phi\right]  }%
{\delta\beta\phi_{\mathbf{r}}}=\rho_{\mu\mathbf{r}}\left[  \phi\right]
,\;\;\text{(GC)}\nonumber
\end{align}
and its useful below to note the elementary result that%
\begin{equation}
\frac{\partial\beta\Omega_{\mu}\left[  \phi\right]  }{\partial\mu
}=-\left\langle N\right\rangle _{\mu}=-\int\rho_{\mu\mathbf{r}}\left[
\phi\right]  d\mathbf{r}\equiv-N_{\mu}, \label{mu}%
\end{equation}
the average number of particles.

Finally, the central actors in the following will be the functionals
\begin{align}
\Lambda_{N}\left[  \phi,\phi_{0}\right]   &  \equiv A_{N}\left[  \phi
_{0}\right]  +k_{B}T\int\widehat{f}_{N}\left[  \phi\right]  \ln\frac
{\widehat{f}_{N}\left[  \phi\right]  }{\widehat{f}_{N}\left[  \phi_{0}\right]
}d\Gamma^{\left(  N\right)  },\;\;\text{(C)}\\
\Lambda_{\mu}\left[  \phi,\phi_{0}\right]   &  \equiv\Omega_{\mu}\left[
\phi_{0}\right]  +k_{B}T\sum_{N=0}^{\infty}\int\widehat{f}_{N\mu}\left[
\phi\right]  \ln\frac{\widehat{f}_{N\mu}\left[  \phi\right]  }{\widehat
{f}_{N\mu}\left[  \phi_{0}\right]  }d\Gamma^{\left(  N\right)  }%
,\;\;\text{(GC)}\nonumber
\end{align}
which can also be written as
\begin{align}
\Lambda_{N}\left[  \phi,\phi_{0}\right]   &  =A_{N}\left[  \phi\right]
+\int\rho_{N\mathbf{r}}\left[  \phi\right]  \left(  \phi_{0,\mathbf{r}}%
-\phi_{\mathbf{r}}\right)  d\mathbf{r},\;\;\text{(C)}\\
\Lambda_{\mu}\left[  \phi,\phi_{0}\right]   &  =\Omega_{\mu}\left[
\phi\right]  +\int\rho_{\mu\mathbf{r}}\left[  \phi\right]  \left(
\phi_{0,\mathbf{r}}-\phi_{\mathbf{r}}\right)  d\mathbf{r}.\;\;\text{(GC)}%
\nonumber
\end{align}

\subsection{Fundamental theorem: relation between fields and densities}

From the Gibbs inequality, one immediately finds that
\begin{align}
\Lambda_{N}\left[  \phi,\phi_{0}\right]   &  \geq A_{N}\left[  \phi
_{0}\right]  ,\;\;\text{(C)}\\
\Lambda\left[  \phi,\phi_{0}\right]   &  \geq\Omega_{\mu}\left[  \phi
_{0}\right]  ,\;\;\text{(GC)}\nonumber
\end{align}
with equality if and only if
\begin{align}
\frac{\widehat{f}_{N}\left[  \phi\right]  }{\widehat{f}_{N}\left[  \phi
_{0}\right]  }  &  =1,\;\;\text{(C)}\\
\frac{\widehat{f}_{N\mu}\left[  \phi\right]  }{\widehat{f}_{N\mu}\left[
\phi_{0}\right]  }  &  =1,\;\;\text{(GC)}\nonumber
\end{align}
for all $\Gamma^{\left(  N\right)  }.$ (Note that this requirement holds up to
a set of measure zero). To understand the meaning of the requirement for
equality, we substitute the explicit expressions for $\widehat{f}_{N}$ and
$\widehat{f}_{N\mu}$ and after rearranging one finds%
\begin{align}
\left\{  \frac{\widehat{f}_{N}\left[  \phi\right]  }{\widehat{f}_{N}\left[
\phi_{0}\right]  }=1\right\}   &  \rightarrow\left\{  \exp\left(  -\beta
\sum_{n=1}^{N}\left(  \phi_{\mathbf{q}_{n}}-\phi_{0\mathbf{q}_{n}}\right)
\right)  =\frac{\int\exp\left(  -\beta\widehat{H}_{N}\left[  \phi\right]
\right)  d\Gamma^{\left(  N\right)  }}{\int\exp\left(  -\beta\widehat{H}%
_{N}\left[  \phi_{0}\right]  \right)  d\Gamma^{\left(  N\right)  }}\right\}
,\;\;\text{(C)}\\
\left\{  \frac{\widehat{f}_{N\mu}\left[  \phi\right]  }{\widehat{f}_{N\mu
}\left[  \phi_{0}\right]  }=1\right\}   &  \rightarrow\left\{  \exp\left(
-\beta\sum_{n=1}^{N}\left(  \phi_{\mathbf{q}_{n}}-\phi_{0\mathbf{q}_{n}%
}\right)  \right)  =\frac{\sum_{N^{\prime}=0}^{\infty}\frac{1}{N^{\prime
}!h^{N^{\prime}D}}\int\exp\left(  -\beta\left(  \widehat{H}_{N^{\prime}%
}\left[  \phi\right]  -\mu N^{\prime}\right)  \right)  d\Gamma^{\left(
N\right)  }}{\sum_{N^{\prime}=0}^{\infty}\frac{1}{N^{\prime}!h^{N^{\prime}D}%
}\int\exp\left(  -\beta\left(  \widehat{H}_{N^{\prime}}\left[  \phi
_{0}\right]  -\mu N^{\prime}\right)  \right)  d\Gamma^{\left(  N\right)  }%
}\right\}  ,\;\;\text{(GC)}\nonumber
\end{align}
The left hand sides of these relations depend on the field at all points in
space while the right hand sides are constants:\ this means that in both cases
the relations can only be satisfied if $\phi_{\mathbf{r}}-\phi_{0\mathbf{r}%
}=c$ for some constant, $c$. Substituting into both sides and using the fact
that $\widehat{H}_{N}\left[  \phi+c\right]  =\widehat{H}_{N}\left[
\phi\right]  +Nc$ then gives
\begin{align}
\left\{  \frac{\widehat{f}_{N}\left[  \phi\right]  }{\widehat{f}_{N}\left[
\phi_{0}\right]  }=1\right\}   &  \rightarrow\left\{  \exp\left(  -\beta
Nc\right)  =\exp\left(  -\beta Nc\right)  \right\}  ,\;\;\text{(C)}\\
\left\{  \frac{\widehat{f}_{N\mu}\left[  \phi\right]  }{\widehat{f}_{N\mu
}\left[  \phi_{0}\right]  }=1\right\}   &  \rightarrow\left\{  \exp\left(
-\beta Nc\right)  =\frac{\sum_{N^{\prime}=0}^{\infty}\frac{1}{N^{\prime
}!h^{N^{\prime}D}}\int\exp\left(  -\beta\left(  \widehat{H}_{N^{\prime}%
}\left[  \phi_{0}\right]  -\left(  \mu-c\right)  N^{\prime}\right)  \right)
d\Gamma^{\left(  N\right)  }}{\sum_{N^{\prime}=0}^{\infty}\frac{1}{N^{\prime
}!h^{N^{\prime}D}}\int\exp\left(  -\beta\left(  \widehat{H}_{N^{\prime}%
}\left[  \phi_{0}\right]  -\mu N^{\prime}\right)  \right)  d\Gamma^{\left(
N\right)  }}\right\}  ,\;\;\text{(GC)}\nonumber
\end{align}
and now the fundamental difference between the ensembles appears:\ the
condition holds in the canonical ensemble for \emph{all values} of the
constant whereas in the grand-canonical ensemble, since the expression must
hold for all $N$ and yet the right hand side is independent of $N$, the only
choice is $c=0$.

Using this information, the result to this point can be summarized as
\begin{align}
\left\{  \phi_{\mathbf{r}}=\phi_{0,\mathbf{r}}+c\right\}  \lor A_{N}\left[
\phi_{0}\right]   &  <A_{N}\left[  \phi\right]  +\int\rho_{N\mathbf{r}}\left[
\phi\right]  \left(  \phi_{0,\mathbf{r}}-\phi_{\mathbf{r}}\right)
d\mathbf{r},\;\;\text{(C)}\\
\left\{  \phi_{\mathbf{r}}=\phi_{0,\mathbf{r}}\right\}  \lor\Omega_{\mu
}\left[  \phi_{0}\right]   &  <\Omega_{\mu}\left[  \phi\right]  +\int\rho
_{\mu\mathbf{r}}\left[  \phi\right]  \left(  \phi_{0,\mathbf{r}}%
-\phi_{\mathbf{r}}\right)  d\mathbf{r},\;\;\text{(GC)}\nonumber
\end{align}
Repeating the derivation but switching the role of the two fields gives
\begin{align}
\left\{  \phi_{\mathbf{r}}=\phi_{0,\mathbf{r}}+c\right\}  \lor A_{N}\left[
\phi\right]   &  <A_{N}\left[  \phi_{0}\right]  +\int\rho_{N\mathbf{r}}\left[
\phi_{0}\right]  \left(  \phi_{\mathbf{r}}-\phi_{0,\mathbf{r}}\right)
d\mathbf{r},\;\;\text{(C)}\\
\left\{  \phi_{\mathbf{r}}=\phi_{0,\mathbf{r}}\right\}  \lor\Omega_{\mu
}\left[  \phi\right]   &  <\Omega_{\mu}\left[  \phi_{0}\right]  +\int\rho
_{\mu\mathbf{r}}\left[  \phi_{0}\right]  \left(  \phi_{\mathbf{r}}%
-\phi_{0,\mathbf{r}}\right)  d\mathbf{r},\;\;\text{(GC)}\nonumber
\end{align}
and adding the two gives
\begin{align}
\left\{  \phi_{\mathbf{r}}=\phi_{0,\mathbf{r}}+c\right\}  \lor0  &
<\int\left(  \rho_{N\mathbf{r}}\left[  \phi\right]  -\rho_{N\mathbf{r}}\left[
\phi_{0}\right]  \right)  \left(  \phi_{0,\mathbf{r}}-\phi_{\mathbf{r}%
}\right)  d\mathbf{r},\;\;\text{(C)}\label{r1}\\
\left\{  \phi_{\mathbf{r}}=\phi_{0,\mathbf{r}}\right\}  \lor0  &  <\int\left(
\rho_{\mu\mathbf{r}}\left[  \phi\right]  -\rho_{\mu\mathbf{r}}\left[  \phi
_{0}\right]  \right)  \left(  \phi_{0,\mathbf{r}}-\phi_{\mathbf{r}}\right)
d\mathbf{r},\;\;\text{(GC)}\nonumber
\end{align}
The import of this result is the conclusion that the densities generated by
two potentials, $\phi$ and $\phi_{0}$, can only be equal if the potentials are
trivially related,
\begin{align}
\rho_{N\mathbf{r}}\left[  \phi\right]   &  = \rho_{N\mathbf{r}}\left[
\phi_{0}\right]  \Rightarrow\phi_{\mathbf{r}}\ = \phi_{0,\mathbf{r}%
}+c,\;\;\text{(C)}\\
\rho_{\mu\mathbf{r}}\left[  \phi\right]   &  = \rho_{\mu\mathbf{r}}\left[
\phi_{0}\right]  \Rightarrow\phi_{\mathbf{r}} = \phi_{0,\mathbf{r}%
},\;\;\text{(GC)}\nonumber
\end{align}
(up to a set of measure zero). It is obvious from the expressions for the
local density, Eq.(\ref{r1}), that the reverse implication holds: each field
(or affine family of fields in the CE) generates a unique local density, so
the final result is
\begin{align}
\rho_{N\mathbf{r}}\left[  \phi\right]   &  = \rho_{N\mathbf{r}}\left[
\phi_{0}\right]  \Leftrightarrow\phi_{\mathbf{r}}\ = \phi_{0,\mathbf{r}%
}+c,\;\;\text{(C)}\\
\rho_{\mu\mathbf{r}}\left[  \phi\right]   &  = \rho_{\mu\mathbf{r}}\left[
\phi_{0}\right]  \Leftrightarrow\phi_{\mathbf{r}} = \phi_{0,\mathbf{r}%
},\;\;\text{(GC)}\nonumber
\end{align}
thus showing that there is a unique mapping between local densities and
fields, in the GCE, or affine families of fields in the CE. One way to
understand the difference between these is that in the canonical ensemble, one
must also supply a gauge condition such to fix the constant such as $\min$
$\phi_{\mathbf{r}}=0$ or $Z_{N}[\phi] = 1$ etc. Given such a condition, the
mapping between fields and densities becomes unique in the canonical ensemble,
just as in the grand-canonical ensemble.

In summary, in the grand canonical ensemble, each field $\phi$ generates a
unique local density $\rho_{\mu\mathbf{r}}\left[  \phi\right]  $ as evidenced
by the explicit formula for the density, Eq.(\ref{explicit}). This means that
if two density fields $\rho_{\mu\mathbf{r}}\left[  \phi_{1}\right]  $ and
$\rho_{\mu\mathbf{r}}\left[  \phi_{2}\right]  $ differ then the fields
$\phi_{1}$ and $\phi_{2}$ cannot be identical at all points. Conversely, two
fields that differ on a set of non-zero measure, $\phi_{1\mathbf{r}}$ and
$\phi_{2\mathbf{r}}$ generate densities which also differ, at least in some
regions of space. What is not proven is that for \textit{any} given density
field $\rho_{\mathbf{r}}$ there exists an external field $\phi_{\mathbf{r}}$
such that $\rho_{\mathbf{r}}=\rho_{\mu\mathbf{r}}\left[  \phi\right]  $. This
is the well-known "v-representability" problem (because the external field is
often called $v$ rather than $\phi$) and in fact, examples will be given below
where this is trivially seen not to be the case. If we let $\mathcal{R}$
denote the set of all local densities that are generated by some field, then
we can say that there is a one-to-one correspondence between fields $\phi$ and
densities $\rho\in\mathcal{R}$. Notice that the set $\mathcal{R}$ is
independent of the chemical potential since it is obviously the case that
\begin{equation}
\rho_{\mu_{1}\mathbf{r}}\left[  \phi\right]  =\rho_{\mu_{2}\mathbf{r}}\left[
\phi+\mu_{2}-\mu_{1}\right]
\end{equation}
so a density that is v-representable at some chemical potential is
representable at any chemical potential.

In the canonical ensemble, we can formulate a similar statement : one can say
that there is a one-to-one correspondence between affine families of fields
$\phi+c$ and densities $\rho\in\mathcal{R}_{N}$. Notice that the density does
not need to be labeled with $N\ $since any density that results from
Eq.(\ref{explicit}) automatically has total number of particles $N$\ :\ this
means that $\int\rho_{\mathbf{r}}d\mathbf{r}=N$ is a necessary condition for
$\rho\in\mathcal{R}_{N}$. Equivalently, if $\mathcal{G}$ is the set of all
potentials satisfying a given gauge condition, then one could say that for a
given there is a one-to-one correspondence between fields $\phi\in\mathcal{G}$
and densities $\rho\in\mathcal{R}_{N}$. Again, there is no proof of
v-representability of any arbitrary $\rho_{\mathbf{r}}$ and, in fact, examples
for which there is no such general representability will be given below
although, as just mentioned, a density with $\int\rho_{\mathbf{r}}%
d\mathbf{r}\neq N$ would be a trivial example.

\subsection{The Helmholtz functional}

Let us write the minimization condition in the form
\begin{align}
A_{N}\left[  \phi_{0}\right]   &  =\min_{\phi\in\mathcal{G}}\left\{
A_{N}\left[  \phi\right]  +\int\rho_{N\mathbf{r}}\left[  \phi\right]  \left(
\phi_{0,\mathbf{r}}-\phi_{\mathbf{r}}\right)  d\mathbf{r}\right\}
,\;\;\text{(C)}\\
\Omega_{\mu}\left[  \phi_{0}\right]   &  =\min_{\phi}\left\{  \Omega_{\mu
}\left[  \phi\right]  +\int\rho_{\mu\mathbf{r}}\left[  \phi\right]  \left(
\phi_{0,\mathbf{r}}-\phi_{\mathbf{r}}\right)  d\mathbf{r}\right\}
.\;\;\text{(GC)}\nonumber
\end{align}
Given the uniqueness of the mappings, one can parameterize the field by
v-representable densities and so get
\begin{align}
A_{N}\left[  \phi_{0}\right]   &  =\min_{\rho\in\mathcal{R}_{N}}\left\{
A_{N}\left[  \phi_{N}\left[  \rho\right]  \right]  +\int\rho_{\mathbf{r}%
}\left(  \phi_{0,\mathbf{r}}-\phi_{N\mathbf{r}}\left[  \rho\right]  \right)
d\mathbf{r}\right\}  ,\;\;\text{(C)}\\
\Omega_{\mu}\left[  \phi_{0}\right]   &  =\min_{\rho\in\mathcal{R}}\left\{
\Omega_{\mu}\left[  \phi_{\mu}\left[  \rho\right]  \right]  +\int
\rho_{\mathbf{r}}\left(  \phi_{0,\mathbf{r}}-\phi_{\mathbf{r}}\left[
\rho\right]  \right)  d\mathbf{r}\right\}  ,\;\;\text{(GC)}\nonumber
\end{align}
or
\begin{align}
A_{N}\left[  \phi_{0}\right]   &  =\min_{\rho\in\mathcal{R}_{N}}\left\{
F_{N}\left[  \rho\right]  +\int\rho_{\mathbf{r}}\phi_{0,\mathbf{r}}%
d\mathbf{r}\right\}  ,\;\;\text{(C)}\\
\Omega_{\mu}\left[  \phi_{0}\right]   &  =\min_{\rho\in\mathcal{R}}\left\{
F\left[  \rho\right]  +\int\rho_{\mathbf{r}}\left(  \phi_{0,\mathbf{r}}%
-\mu\right)  d\mathbf{r}\right\}  ,\;\;\text{(GC)}\nonumber
\end{align}
with%
\begin{align}
F_{N}\left[  \rho\right]   &  \equiv A_{N}\left[  \phi_{N}\left[  \rho\right]
\right]  -\int\rho_{\mathbf{r}}\phi_{N\mathbf{r}}\left[  \rho\right]
d\mathbf{r},\;\;\text{(C)}\label{f}\\
F\left[  \rho\right]   &  \equiv\Omega_{\mu}\left[  \phi_{\mu}\left[
\rho\right]  \right]  -\int\rho_{\mathbf{r}}\left(  \phi_{\mu\mathbf{r}%
}\left[  \rho\right]  -\mu\right)  d\mathbf{r}.\;\;\text{(GC)}\nonumber
\end{align}
Note that in the grand-canonical ensemble, the so-called "Helmholtz"
functional $F\left[  \rho\right]  $ does not depend on the chemical potential
as is easily verified from
\begin{align}
\left.  \frac{\partial F\left[  \rho\right]  }{\partial\mu}\right\vert
_{\rho\mu}  &  =\left.  \frac{\partial\Omega_{\mu}\left[  \phi_{\mu}\left[
\rho\right]  \right]  }{\partial\mu}\right\vert _{\rho}-\int\rho_{\mathbf{r}%
}\left(  \frac{\partial\phi_{\mu\mathbf{r}}\left[  \rho\right]  }{\partial\mu
}-1\right)  d\mathbf{r}\\
&  =\left.  \frac{\partial\Omega_{\mu}\left[  \phi\right]  }{\partial\mu
}\right\vert _{\phi_{\mu}\left[  \rho\right]  }+\int\left.  \frac{\delta
\Omega_{\mu}\left[  \phi\right]  }{\delta\phi_{\mathbf{r}}}\right\vert
_{\phi_{\mu}\left[  \rho\right]  }\frac{\partial\phi_{\mu\mathbf{r}}\left[
\rho\right]  }{\partial\mu}d\mathbf{r}-\int\rho_{\mathbf{r}}\left(
\frac{\partial\phi_{\mu\mathbf{r}}\left[  \rho\right]  }{\partial\mu
}-1\right)  d\mathbf{r}\nonumber\\
&  =-N_{\mu}\left[  \phi_{\mu}\left[  \rho\right]  \right]  +\int
\rho_{\mathbf{r}}\frac{\partial\phi_{\mu\mathbf{r}}\left[  \rho\right]
}{\partial\mu}d\mathbf{r}-\int\rho_{\mathbf{r}}\left(  \frac{\partial\phi
_{\mu\mathbf{r}}\left[  \rho\right]  }{\partial\mu}-1\right)  d\mathbf{r}%
\nonumber\\
&  =-\int\rho_{\mathbf{r}}\left[  \phi_{\mu}\left[  \rho\right]  \right]
d\mathbf{r}+\int\rho_{\mathbf{r}}d\mathbf{r}\nonumber\\
&  =0\nonumber
\end{align}
since by definition $\rho_{\mathbf{r}}\left[  \phi_{\mu}\left[  \rho\right]
\right]  =\rho_{\mathbf{r}}$. This gives the central result that the canonical
(grand canonical) free energy is obtained by minimizing the functional%
\begin{align}
\Lambda_{N}\left[  \rho;\phi_{0}\right]   &  =F_{N}\left[  \rho\right]
+\int\rho_{\mathbf{r}}\phi_{0,\mathbf{r}}d\mathbf{r},\;\;\text{(C)}\\
\Lambda_{\mu}\left[  \rho;\phi_{0}\right]   &  =F\left[  \rho\right]
+\int\rho_{\mathbf{r}}\left(  \phi_{0,\mathbf{r}}-\mu\right)  d\mathbf{r}%
,\;\;\text{(GC)}\nonumber
\end{align}
over the density with the minimizing density being $\rho_{N}\left[  \phi
_{0}\right]  $ and $\rho_{\mu}\left[  \phi_{0}\right]  $, respectively, and
with the values of the functionals $\Lambda_{N}$, respectively $\Lambda_{\mu}%
$, at that minimizing density being the free energies for the field $\phi_{0}%
$. Thus, aside from the irrelevant technicality of the gauge condition, the
main formal difference between DFT in the canonical and grand canonical
ensemble is the definition of the v-representable densities. In particular, in
the canonical ensemble, the minimization with respect to densities must
obviously respect the canonical condition that the particle number is fixed.
The Helmholtz functionals, $F\left[  \rho\right]  $ and $F_{N}\left[
\rho\right]  $, are universal in the sense that they depend only on the
interaction potential and on the temperature: knowing these, the free energy
for any inhomogeneity-inducing external field, $\phi\left(  \mathbf{r}\right)
$, can be obtained via minimization with respect to the one-body density.

\section{Exact results}

\subsection{Eliminating the momenta}

All exact results begin with the evaluation of the partition function and
density in terms of the field. If the Hamiltonian is written as
\begin{equation}
\widehat{H}_{N}\left[  \phi\right]  =\sum_{i=1}^{N}\frac{p_{i}^{2}}%
{2m}+\widehat{U}_{N}+\sum_{i=1}^{N}\phi_{\mathbf{q}_{i}}%
\end{equation}
then the partition functions become%
\begin{align}
Z_{N}\left[  \phi\right]   &  =\frac{1}{N!}\Lambda_{T}^{-ND}\int\exp\left(
-\beta\widehat{U}_{N\mathbf{q}^{\left(  N\right)  }}-\sum_{i=1}^{N}\beta
\phi_{\mathbf{q}_{i}}\right)  d\mathbf{q}^{\left(  N\right)  },\;\;\text{(C)}%
\\
\Xi_{\mu}\left[  \phi\right]   &  =\sum_{N=0}^{\infty}\frac{1}{N!}\Lambda
_{T}^{-ND}\int\exp\left(  -\beta\widehat{U}_{N\mathbf{q}^{\left(  N\right)  }%
}-\sum_{i=1}^{N}\beta\phi_{\mathbf{q}_{i}}+\beta\mu N\right)  d\mathbf{q}%
^{\left(  N\right)  },\;\text{(GC)}\nonumber
\end{align}
where the thermal wavelength is%
\begin{equation}
\Lambda_{T}=\frac{h}{\sqrt{2\pi mk_{B}T}}.
\end{equation}

\subsection{The ideal gas}

The first example to illustrate the differences between the ensembles is the
ideal gas for which the interaction potential is zero so%
\begin{align}
Z_{N}\left[  \phi\right]   &  =\frac{1}{N!}\left(  \Lambda_{T}^{-D}\int
\exp\left(  -\beta\phi_{\mathbf{q}}\right)  d\mathbf{q}\right)  ^{N}%
,\;\;\text{(C)}\\
\Xi_{\mu}\left[  \phi\right]   &  =\exp\left(  \exp\left(  \beta\mu\right)
\Lambda_{T}^{D}\int\exp\left(  -\beta\phi_{\mathbf{q}}\right)  d\mathbf{q}%
\right)  ,\;\text{(GC)}\nonumber
\end{align}
and the free energies are
\begin{align}
\beta A_{N}\left[  \phi\right]   &  =-\ln\frac{1}{N!}\left(  \Lambda_{T}%
^{-D}\int\exp\left(  -\beta\phi_{\mathbf{q}}\right)  d\mathbf{q}\right)
^{N},\;\;\text{(C)}\\
\beta\Omega_{\mu}\left[  \phi\right]   &  =-\exp\left(  \beta\mu\right)
\Lambda_{T}^{D}\int\exp\left(  -\beta\phi_{\mathbf{q}}\right)  d\mathbf{q}%
,\;\text{(GC)}\nonumber
\end{align}
which, via Eq.(\ref{D}), imply the local densities
\begin{align}
\rho_{N\mathbf{r}}  &  =N\frac{\exp\left(  -\beta\phi_{\mathbf{r}}\right)
}{\int\exp\left(  -\beta\phi_{\mathbf{q}}\right)  d\mathbf{q}},\;\;\text{(C)}%
\\
\rho_{\mu\mathbf{r}}  &  =\Lambda_{T}^{D}\exp\left(  \beta\mu\right)
\exp\left(  -\beta\phi_{\mathbf{q}}\right)  .\;\text{(GC)}\nonumber
\end{align}
The next step is to invert this relation. In the canonical case, it is clear
that $\exp\left(  -\beta\phi_{\mathbf{r}}\right)  \propto\Lambda^{D}%
\rho_{N\mathbf{r}}$ but there is no way to determine the proportionality
constant without specifying the gauge. This is of no concern as we simply
write
\begin{align}
\beta\phi_{\mathbf{r}}\left[  \rho_{\mathbf{r}}\right]   &  =\beta
c-\ln\left(  \Lambda_{T}^{D}\rho_{\mathbf{r}}\right)  ,\;\;\text{(C)}\\
\beta\phi_{\mu\mathbf{r}}\left[  \rho_{\mathbf{r}}\right]   &  =\beta\mu
-\ln\left(  \Lambda_{T}^{D}\rho_{\mathbf{r}}\right)  ,\;\text{(GC)}\nonumber
\end{align}
where $c$ is arbitrary. The partition functions can then be expressed in terms
of the density as
\begin{align}
Z_{N}\left[  \phi\left[  \rho\right]  \right]   &  =\frac{1}{N!}\left(
\exp\left(  -\beta c\right)  N\right)  ^{N},\;\;\text{(C)}\\
\Xi_{\mu}\left[  \phi\left[  \rho\right]  \right]   &  =\exp\left(  \int
\rho_{\mathbf{r}}d\mathbf{q}\right)  ,\;\text{(GC)}\nonumber
\end{align}
giving the free energies
\begin{align}
\beta A_{N}\left[  \phi\left[  \rho\right]  \right]   &  =-\ln\left(  \frac
{1}{N!}\left(  e^{-\beta c}N\right)  ^{N}\right)  ,\;\;\text{(C)}\\
\beta\Omega_{\mu}\left[  \phi_{\mu}\left[  \rho\right]  \right]   &
=-\int\rho_{\mathbf{q}}d\mathbf{q}.\;\text{(GC)}\nonumber
\end{align}
Substituting into Eq.(\ref{f}) gives the Helmholtz functionals
\begin{align}
F_{N}\left[  \rho\right]   &  =-\ln\left(  \frac{1}{N!}\left(  e^{-\beta
c}N\right)  ^{N}\right)  -\int\rho_{\mathbf{r}}\left(  \beta c-\ln\Lambda
^{D}\rho_{N\mathbf{r}}\right)  d\mathbf{r},\;\;\text{(C)}\\
F\left[  \rho\right]   &  =-\int\rho_{\mathbf{r}}d\mathbf{r}+\int
\rho_{\mathbf{r}}\ln\left(  \Lambda^{D}\rho_{\mathbf{r}}\right)
d\mathbf{r},\;\;\text{(GC)}\nonumber
\end{align}
which can be written as
\begin{align}
F_{N}\left[  \rho_{N}\right]   &  =\int\rho_{\mathbf{r}}\ln\left(  \Lambda
^{D}\rho_{\mathbf{r}}\right)  d\mathbf{r}-\ln\frac{N^{N}}{N!},\;\;\text{(C)}%
\label{idealgas}\\
F\left[  \rho\right]   &  =\int\left\{  \rho_{\mathbf{r}}\ln\left(
\Lambda^{D}\rho_{\mathbf{r}}\right)  -\rho_{\mathbf{r}}\right\}
d\mathbf{r}.\;\;\text{(GC)}\nonumber
\end{align}
Note that the gauge constant does not appear in the final result for the
canonical ensemble. Using Sterling's approximation, one sees that in the limit
of large $N$,
\begin{equation}
\ln\frac{N^{N}}{N!}=N\left(  1+O\left(  \frac{\ln\left(  N\right)  }%
{N}\right)  \right)
\end{equation}
so one can write%
\begin{equation}
F_{N}\left[  \rho\right]  =\int\left\{  \rho_{\mathbf{r}}\ln\left(
\Lambda^{D}\rho_{\mathbf{r}}\right)  -\rho_{\mathbf{r}}\left(  1+O\left(
\frac{\ln N}{N}\right)  \right)  \right\}  d\mathbf{r},\;\;\text{(C)}%
\end{equation}
showing that the functional becomes the same as that for the grand canonical
ensemble in the limit of large $N$. This reproduces the result previously
given by White et al\cite{W1,W2}.

\subsection{Hard particles in a restricted geometry}

An example that has played an important role in recent years\cite{lutsko2020}
is that of a system of identical hard particles confined to a set of cavities
each of which is large enough to hold one, but not two, of the particles.
A\ further complication is that the cavities may overlap in such a way that if
one is filled, then one or more of the others is partially filled and so
blocked. In the grand canonical ensemble, this is quite non trivial,
especially in the case of overlapping cavities, since each may hold either
zero or one particles but exact results are nevertheless possible since the
sum over particle number is restricted by the number of cavities. Here the
functionals for linear chains of one or more such cavities which overlap in
such a way that if one cavity is filled, then its neighbors cannot be occupied
(see Figure). In the following discussion, the center of the i-th cavity will
be $\mathbf{s}_{i}$, and we define the dimensionless quantities%
\begin{align}
e_{i}  &  =\Lambda_{T}^{-D}\int_{V_{i}}\exp\left(  -\beta\phi_{\mathbf{r}%
}\right)  d\mathbf{r}\\
N_{i}  &  =\int_{V_{i}}\rho\left(  \mathbf{r}\right)  d\mathbf{r}\nonumber
\end{align}
where the integrals are restricted to the volume accessible to the center of
mass of a particle and the second quantity is the average number of particles
in the i-th cavity. When considering the grand canonical ensemble, the
definition of $e_{i}$ will be modified with the replacement $\phi_{\mathbf{r}%
}\rightarrow\phi_{\mathbf{r}}-\mu$.

\begin{figure}
[htp!]\includegraphics[width=0.40\linewidth]{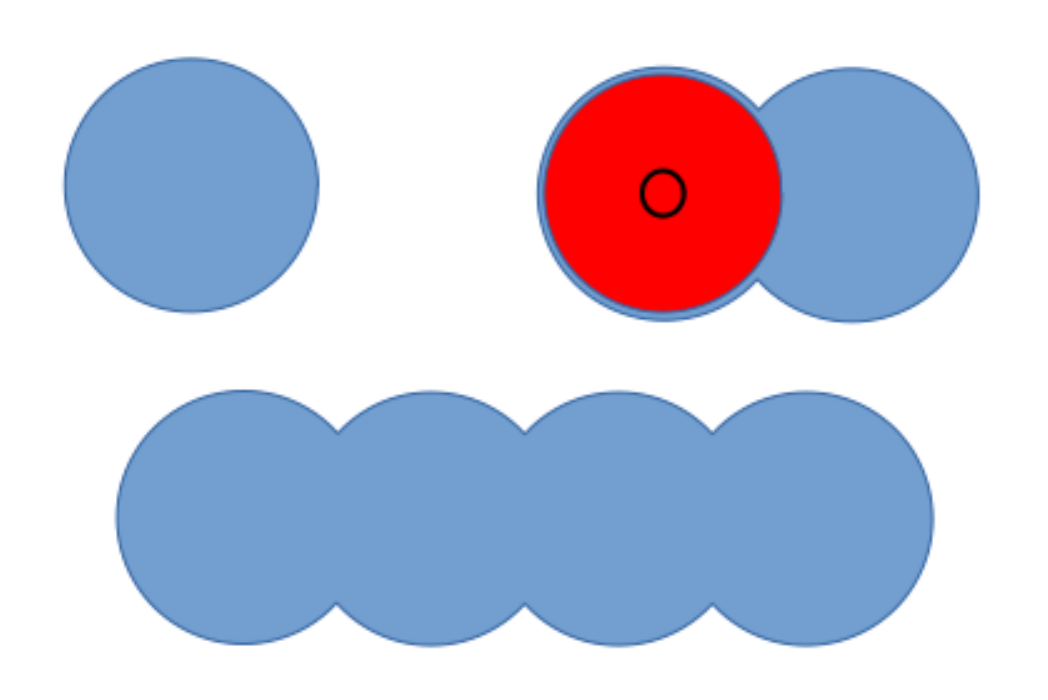}
\caption{Chains of one, two and four overlapping cavities. The chain of length two shows a hard disk and the small black circle is the volume accessible to its center in the first cavity.}
\label{chains}
\end{figure}

\subsubsection{One particle in a chain of $M$ cavities}

To see what happens in the canonical ensemble, consider the case of a chain of
\ $M$ such cavities in $D$ dimensions. The case of $N_{c}=1$ is referred to as
a "zero-dimensional" system in the limit that the cavity is just large enough
to hold a single particle. Elementary evaluations lead to
\begin{align}
Z_{1}\left[  \phi\right]   &  =\sum_{i=1}^{M}e_{i}\\
\rho_{\mathbf{r}} &  =\frac{e^{-\beta\phi_{\mathbf{r}}}}{Z_{1}\left[
\phi\right]  }\sum_{i=1}^{M}\delta\left(  \mathbf{r}\in V_{1}\right)
\nonumber
\end{align}
so that
\begin{equation}
e^{-\beta\phi_{\mathbf{r}}}=Z_{1}\left[  \phi\right]  \Lambda_{T}^{D}%
\rho_{\mathbf{r}}%
\end{equation}
and%
\begin{align}
\beta F_{1}\left[  \rho\right]   &  =-\ln\left(  Z_{1}\left[  \phi\left[
\rho\right]  \right]  \right)  +\int\rho_{\mathbf{r}}\ln\left(  Z_{1}\left[
\phi\left[  \rho\right]  \right]  \Lambda_{T}^{D}\rho_{\mathbf{r}}\right)
d\mathbf{r}\\
&  =\int\rho_{\mathbf{r}}\ln\left(  \Lambda_{T}^{D}\rho_{\mathbf{r}}\right)
d\mathbf{r}\nonumber
\end{align}
which is the ideal-gas result, as one would guess. For comparison, the
grand-canonical functional is
\begin{equation}
\beta F\left[  \rho\right]  =\beta F^{\left(  \text{id}\right)  }\left[
\rho\right]  +\Phi\left(  \sum_{i=1}^{M}N_{1}\right)  \label{gce_1}%
\end{equation}
with $\Phi\left(  x\right)  =\left(  1-x\right)  \ln\left(  1-x\right)  +x$
(see, e.g. Ref.\cite{lutsko2020}). The excess functional - the correction to
the ideal gas - is nonzero solely due to the fluctuations in particle number.

\subsubsection{Two particles in a chain of three cavities}

If the three cavities do not overlap the functional can be guessed based on
the preceding results. In the new case of a chain of overlapping cavities the
results in both ensembles are non-trivial. We first consider the grand
canonical ensemble for which the partition function is
\begin{equation}
\Xi_{\mu}\left[  \phi\right]  =1+e_{1}+e_{2}+e_{3}+e_{1}e_{3}%
\end{equation}
giving the density%
\begin{equation}
\rho_{\mu\mathbf{r}}=e^{-\beta\left(  \phi_{\mathbf{r}}-\mu\right)  }%
\frac{\left(  1+e_{3}\right)  \delta\left(  \mathbf{r}\in V_{1}\right)
+\delta\left(  \mathbf{r}\in V_{2}\right)  +\left(  1+e_{1}\right)
\delta\left(  \mathbf{r}\in V_{3}\right)  }{\Xi_{\mu}\left[  \phi\right]  }%
\end{equation}
This is integrated over each cavity to get
\begin{equation}
N_{1}=\frac{1+e_{3}}{\Xi_{\mu}\left[  \phi\right]  }e_{1},\;N_{2}=\frac{1}%
{\Xi_{\mu}\left[  \phi\right]  }e_{2},\;N_{3}=\frac{1+e_{1}}{\Xi_{\mu}\left[
\phi\right]  }e_{3}%
\end{equation}
and from this system we find
\begin{align}
e_{1}  &  =\frac{N_{1}}{1-N_{1}-N_{2}},e_{2}=\frac{N_{2}-N_{2}^{2}}{\left(
N_{2}+N_{3}-1\right)  \left(  N_{1}+N_{2}-1\right)  },e_{3}=\frac{N_{3}%
}{1-N_{2}-N_{3}}\\
\Xi &  =\frac{1-N_{2}}{\left(  N_{1}+N_{2}-1\right)  \left(  N_{2}%
+N_{3}-1\right)  }\nonumber
\end{align}
and%
\begin{equation}
e^{-\beta\left(  \phi_{\mathbf{r}}-\mu\right)  }=\Xi_{\mu}\left[  \phi\right]
\rho_{\mu\mathbf{r}}\left\{  \left(  1-\frac{N_{3}}{1-N_{2}}\right)
\delta\left(  \mathbf{r}\in V_{1}\right)  +\delta\left(  \mathbf{r}\in
V_{2}\right)  +\left(  1-\frac{N_{1}}{1-N_{2}}\right)  \delta\left(
\mathbf{r}\in V_{3}\right)  \right\}
\end{equation}
so that the grand canonical functional is
\begin{equation}
\beta F\left[  \rho\right]  =\int\rho_{\mathbf{r}}\ln\rho_{\mathbf{r}%
}d\mathbf{r}-N_{1}\ln\left(  \frac{N_{3}}{1-N_{2}}\right)  -N_{3}\ln\left(
\frac{N_{1}}{1-N_{2}}\right)
\end{equation}

In the canonical ensemble, one particle is an ideal gas so we turn to the case
of two particles. Here
\begin{align}
Z_{2}\left[  \phi\right]   &  =e_{1}e_{3}\\
\rho_{2\mathbf{r}}  &  =\frac{e^{-\beta\phi_{\mathbf{r}}}}{e_{1}}\delta\left(
\mathbf{r}\in V_{1}\right)  +\frac{e^{-\beta\phi_{\mathbf{r}}}}{e_{3}}%
\delta\left(  \mathbf{r}\in V_{3}\right) \nonumber
\end{align}
so%
\begin{equation}
e^{-\beta\phi_{\mathbf{r}}}=e_{1}\rho_{2\mathbf{r}}\delta\left(  \mathbf{r}\in
V_{1}\right)  +e_{3}\rho_{2\mathbf{r}}\delta\left(  \mathbf{r}\in
V_{3}\right)
\end{equation}
and it follows that
\begin{equation}
\beta F_{2}\left[  \rho\right]  =\int_{V_{1}}\rho_{\mathbf{r}}\ln
\rho_{\mathbf{r}}d\mathbf{r+}\int_{V_{3}}\rho_{\mathbf{r}}\ln\rho_{\mathbf{r}%
}d\mathbf{r}%
\end{equation}
which is not an ideal gas unless the potential happens to forbid occupancy
(i.e. to be infinite) in the middle cavity. Note that there are several a
priori constraints on the density:\ $N_{1}=N_{3}=1$ and $N_{2}=0$.

\subsubsection{Two particles in a chain of four cavities}

For two particles in a chain of four cavities, the canonical partition
function is%

\begin{equation}
Z_{2}\left[  \phi\right]  =e_{1}e_{3}+e_{1}e_{4}+e_{2}e_{4}%
\end{equation}
and repeating the usual steps one finds the field
\begin{equation}
Z_{2}\left[  \phi\right]  e^{-\beta\phi_{\mathbf{r}}}=\frac{1}{e_{3}+e_{4}%
}\rho_{2\mathbf{r}}\delta\left(  \mathbf{r}\in V_{1}\right)  +\frac{1}{e_{4}%
}\rho_{2\mathbf{r}}\delta\left(  \mathbf{r}\in V_{2}\right)  +\frac{1}{e_{3}%
}\rho_{2\mathbf{r}}\delta\left(  \mathbf{r}\in V_{3}\right)  +\frac{1}%
{e_{1}+e_{2}}\rho_{2\mathbf{r}}\delta\left(  \mathbf{r}\in V_{4}\right)
\end{equation}
giving the Helmholtz functional%
\begin{align}
F_{2}\left[  \rho\right]   &  =\int\rho_{2\mathbf{r}}\ln\rho_{2\mathbf{r}%
}d\mathbf{r}+N_{1}\ln\frac{1}{e_{3}+e_{4}}+N_{2}\ln\frac{1}{e_{4}}+N_{3}%
\ln\frac{1}{e_{1}}+N_{4}\ln\frac{1}{e_{1}+e_{2}}\\
&  +\ln\left(  e_{1}e_{3}+e_{1}e_{4}+e_{2}e_{4}\right) \nonumber
\end{align}
with the constants determined from
\begin{align}
Z_{2}\left[  \phi\right]  N_{1}  &  =\left(  e_{3}+e_{4}\right)  e_{1}\\
Z_{2}\left[  \phi\right]  N_{2}  &  =e_{2}e_{4}\nonumber\\
Z_{2}\left[  \phi\right]  N_{3}  &  =e_{1}e_{3}\nonumber\\
Z_{2}\left[  \phi\right]  N_{4}  &  =\left(  e_{1}+e_{2}\right)
e_{4}\nonumber
\end{align}
The physical requirements that one particle be in one of the first two
cavities and the second in one of the last are reflected in the degeneracy
$N_{1}+N_{2}=N_{3}+N_{4}=1$, so there are only two independent equations
giving, e.g.
\begin{align}
\frac{N_{1}}{N_{3}}  &  =\frac{e_{4}}{e_{3}}+1\rightarrow e_{3}=e_{4}\left(
\frac{N_{1}}{N_{3}}-1\right)  ^{-1}\\
\frac{N_{4}}{N_{2}}  &  =\frac{e_{1}}{e_{2}}+1\rightarrow e_{2}=e_{1}\left(
\frac{N_{4}}{N_{2}}-1\right)  ^{-1}\nonumber
\end{align}
and finally%
\begin{align}
F_{2}\left[  \rho\right]   &  =\int\rho_{2\mathbf{r}}\ln\rho_{2\mathbf{r}%
}d\mathbf{r}+N_{1}\ln\left(  1-\frac{N_{3}}{N_{1}}\right)  +N_{4}\ln\left(
1-\frac{N_{2}}{N_{4}}\right)  -\ln\left(  \frac{N_{1}N_{4}-N_{2}N_{3}}{\left(
N_{1}-N_{3}\right)  \left(  N_{4}-N_{2}\right)  }\right) \\
&  =\int\rho_{2\mathbf{r}}\ln\rho_{2\mathbf{r}}d\mathbf{r}+\left(  N_{1}%
+N_{4}+1\right)  \ln\left(  N_{1}-N_{3}\right)  -N_{1}\ln N_{1}-N_{4}\ln
N_{4}\nonumber
\end{align}

For comparison, the grand canonical ensemble gives%
\begin{equation}
\Xi\left[  \phi\right]  =1+e_{1}+e_{2}+e_{3}+e_{4}+e_{1}e_{3}+e_{1}e_{4}%
+e_{2}e_{4}%
\end{equation}
and
\begin{align}
e_{1}  &  =\frac{N_{1}}{1-N_{1}-N_{2}}\\
e_{2}  &  =\frac{N_{2}\left(  N_{2}-1\right)  }{\left(  1-N_{2}-N_{3}\right)
\left(  N_{1}+N_{2}-1\right)  }\nonumber\\
e_{3}  &  =\frac{N_{3}\left(  N_{3}-1\right)  }{\left(  1-N_{2}-N_{3}\right)
\left(  N_{3}+N_{4}-1\right)  }\nonumber\\
e_{4}  &  =\frac{N_{4}}{1-N_{3}-N_{4}}\nonumber
\end{align}
yielding
\begin{align}
\beta F\left[  \rho\right]   &  \equiv\int\rho_{\mathbf{r}}\ln\left(
\rho_{\mathbf{r}}\right)  d\mathbf{r}+\left(  1-N_{1}-N_{3}-N_{4}\right)
\ln\left(  1-N_{2}\allowbreak\right) \\
&  +\left(  1\allowbreak-N_{1}-N_{2}-N_{4}\right)  \ln\left(  1-N_{3}%
\allowbreak\right) \nonumber\\
&  -\left(  1-N_{1}-N_{4}\right)  \ln\left(  1-N_{2}-N_{3}\right) \nonumber\\
&  -\left(  1-N_{3}-N_{4}\right)  \ln\left(  1-N_{1}-N_{2}\allowbreak\right)
\nonumber\\
&  -\left(  1-N_{1}-N_{2}\right)  \ln\left(  1-N_{3}-N_{4}\right) \nonumber
\end{align}

\subsubsection{Comments on v-representability}

Notice that all of these results imply certain limits on v-representability.
For example, in the case of a single cavity, in the grand-canonical ensemble
the average particle number is restricted to be $N_{1} < 1$ (see Eq.
\ref{gce_1}). In the case of a chain of four cavities, in the canonical
ensemble one has that $N_{1}+N_{2} = N_{3}+N_{4} = 1$ since there must be two
particles and since adjacent cavities cannot be simultaneously occupied. Any
density violating these constraints cannot be generated by a field. Similarly,
in the grand-canonical ensemble, it must be that $N_{1}+N_{2} < 1$ and $N_{3}
+ N_{4} < 1$ for similar reasons: when there are zero particles, both sums are
zero, when there is one particle neither sum can be greater than one and for
two particles the canonical condition holds. So, the weighted average of these
giving the grand-canonical result is necessarily less than one and any density
violating this is not v-representable.

\section{Hard rods in a cavity}

Two classes of exactly solvable models have played important roles in the
development of modern cDFT in the grand-canonical ensemble. The first is that
of hard-spheres in one dimension, also known as hard rods. The exact Helmholtz
functional for hard rods was found by Percus in 1976 and will be given below.
As of now, no equivalent result is known for the canonical ensemble. Attempts
to generalize Percus' result to higher dimensions eventually led to the
development of Fundamental Measure Theory (FMT)\ which is widely viewed as the
most sophisticated model functional. The development of FMT was further guided
by the second class of exact models, already discussed above, which are hard
particles in small cavities.

\subsection{Grand-canonical}

In the grand-canonical ensemble, the exact Helmholtz functional for the case
of a single species of hard-rods of length $\sigma$ can be written as
\begin{equation}
\label{Percus1}F\left[  \rho\right]  =\int_{-\infty}^{\infty}\rho_{x}\left(
\ln\rho_{x}-1\right)  dx-\int_{-\infty}^{\infty}s_{x}\left[  \rho\right]
\ln\left(  1-\eta_{x}\left[  \rho\right]  \right)  dx
\end{equation}
with%
\begin{align}
\label{Percus2}\eta_{x}\left[  \rho\right]   &  =1-\int_{x-\frac{\sigma}{2}%
}^{x+\frac{\sigma}{2}}\rho_{y}dy\\
s_{x}\left[  \rho\right]   &  =\frac{1}{2}\left(  \rho_{x-\frac{\sigma}{2}%
}+\rho_{x+\frac{\sigma}{2}}\right)  .\nonumber
\end{align}
If there is a hard wall at $x=0$ and at $x=L$ that means that the center of a
hard rod is confined to the domain $\left[  \frac{\sigma}{2},L-\frac{\sigma
}{2}\right]  $ and so the external field is infinite and the density $\rho
_{x}=0$ outside this domain. Consider the case that $L<2\sigma$ so that the
cavity can only hold a single rod. Some elements of the grand canonical
ensemble will have zero rods and some will have one rod so the average total
particle number is between zero and one. In general, $\eta_{x}\left[
\rho\right]  $ will therefore always be between zero and one. Furthermore, if
$x-\frac{\sigma}{2}<\frac{\sigma}{2}$, so that $\rho_{x-\frac{\sigma}{2}}=0$
then $x<\sigma$ and so $x+\frac{\sigma}{2}<\frac{3\sigma}{2}<L-\frac{\sigma
}{2}$ so $s_{x}\left[  \rho\right]  $gives (in general) a nonzero
contribution. This is all to say that the non-ideal gas part of $F\left[
\rho\right]  $ contributes, as expected. Nothing conceptually changes as the
size of the cavity increases except that the maximum value of the average
number of particles.

\subsection{Canonical Ensemble}

As noted above, a single particle in a cavity is just an ideal gas, so the
simplest nontrivial example would involve two particles. In the following, it
is assumed that the length of the cavity is in the range $3\sigma<L<4\sigma$.
The reason for not directly considering the possibility $2\sigma<L<3\sigma$ is
that it gives rise to mathematical difficulties that will be discussed below.

\subsubsection{The local density}

The partition function for the system is
\begin{equation}
Z_{2}\left[  \phi\right]  =\frac{1}{2!}\Lambda_{T}^{2}\int_{\frac{\sigma}{2}%
}^{\frac{5\sigma}{2}+\Delta}\int_{\frac{\sigma}{2}}^{\frac{5\sigma}{2}+\Delta
}e^{-\beta\phi_{y_{1}}}e^{-\beta\phi_{y_{2}}}\Theta\left(  \left\vert
y_{1}-y_{2}\right\vert -\sigma\right)  dy_{1}dy_{2}%
\end{equation}
where $L=3\sigma+\Delta$ and so $0<\Delta<\sigma$ and the step function
$\Theta\left(  z\right)  =1$ for $z>0$ and zero otherwise. The local density
is
\begin{equation}
Z_{2}\left[  \phi\right]  \Lambda_{T}^{2}\rho_{x}=\Theta\left(  \frac{5\sigma
}{2}+\Delta-x\right)  \Theta\left(  x-\frac{\sigma}{2}\right)  e^{-\beta
\phi_{x}}\int_{\frac{\sigma}{2}}^{\frac{5\sigma}{2}+\Delta}e^{-\beta\phi_{y}%
}\Theta\left(  \left\vert x-y\right\vert -\sigma\right)  dy
\end{equation}
which can be written more explicitly as
\begin{align}
Z_{2}\left[  \phi\right]  \Lambda_{T}^{2}\rho_{x}  &  =\Theta\left(
\frac{5\sigma}{2}+\Delta-x\right)  \Theta\left(  x-\frac{3\sigma}{2}\right)
e^{-\beta\phi_{x}}\int_{\frac{\sigma}{2}}^{x-\sigma}e^{-\beta\phi_{y}}dy\\
&  +\Theta\left(  \frac{3\sigma}{2}+\Delta-x\right)  \Theta\left(
x-\frac{\sigma}{2}\right)  e^{-\beta\phi_{x}}\int_{x+\sigma}^{\frac{5\sigma
}{2}+\Delta}e^{-\beta\phi_{y}}dy\nonumber
\end{align}
or, even more explicitly, the density is zero except for
\begin{subequations}
\label{d1}%
\begin{align}
\Lambda_{T}^{2}Z^{\left(  2\right)  }\left[  \phi\right]  \rho_{x}  &
=e^{-\beta\phi_{x}}\int_{x+\sigma}^{\frac{5\sigma}{2}+\Delta}e^{-\beta\phi
_{y}}dy,\;\;\;\;\;\;\;\;\;\;\;\;\;\;\;\;\;\;\;\;\;\;\frac{\sigma}{2}%
<x<\frac{3\sigma}{2}\label{da}\\
&  =e^{-\beta\phi_{x}}\int_{\frac{\sigma}{2}}^{x-\sigma}e^{-\beta\phi_{y}%
}dy+e^{-\beta\phi_{x}}\int_{x+\sigma}^{\frac{5\sigma}{2}+\Delta}e^{-\beta
\phi_{y}}dy,\;\frac{3\sigma}{2}<x<\frac{3\sigma}{2}+\Delta\label{db}\\
&  =e^{-\beta\phi_{x}}\int_{\frac{\sigma}{2}}^{x-\sigma}e^{-\beta\phi_{y}%
}dy,\;\;\;\;\;\;\;\;\;\;\;\;\;\;\;\;\;\;\;\;\;\;\;\;\frac{3\sigma}{2}%
+\Delta<x<\frac{5\sigma}{2}+\Delta. \label{dc}%
\end{align}
One sees immediately that the function $\rho_{x}e^{\beta\phi_{x}}$ is
continuous although its first derivative is not and in fact satisfies the jump
conditions%
\end{subequations}
\begin{align}
\lim_{\epsilon\rightarrow0}\left(  \frac{d}{dx}\rho_{x}e^{\beta\phi_{x}%
}\right)  _{\frac{3\sigma}{2}+\epsilon}  &  =\lim_{\epsilon\rightarrow
0}\left(  \frac{d}{dx}\rho_{x}e^{\beta\phi_{x}}\right)  _{\frac{3\sigma}%
{2}-\epsilon}+\frac{1}{\Lambda_{T}^{2}Z^{\left(  2\right)  }\left[
\phi\right]  }e^{-\beta\phi_{\frac{\sigma}{2}}}\label{jump}\\
\lim_{\epsilon\rightarrow0}\left(  \frac{d}{dx}\rho_{x}e^{\beta\phi_{x}%
}\right)  _{\frac{3\sigma}{2}+\Delta+\epsilon}  &  =\lim_{\epsilon
\rightarrow0}\left(  \frac{d}{dx}\rho_{x}e^{\beta\phi_{x}}\right)
_{\frac{3\sigma}{2}+\Delta-\epsilon}+\frac{1}{\Lambda_{T}^{2}Z^{\left(
2\right)  }\left[  \phi\right]  }e^{-\beta\phi_{\frac{5\sigma}{2}+\Delta}%
}\nonumber
\end{align}
The density also obeys the relation%
\begin{equation}
\rho_{x}e^{\beta\phi_{x}}+\rho_{x+2\sigma}e^{\beta\phi_{x+2\sigma}}=\frac
{1}{\Lambda_{T}^{2}Z^{\left(  2\right)  }\left[  \phi\right]  }\int
_{\frac{\sigma}{2}}^{\frac{5\sigma}{2}+\Delta}e^{-\beta\phi_{y}}dy\equiv
A,\;\;\;\;\;\frac{\sigma}{2}<x<\frac{\sigma}{2}+\Delta\label{dual}%
\end{equation}
which is easily verified by substituting the appropriate expressions for the
density from Eq.(\ref{d1}). This will be referred to as the "duality" relation
since it tells us that the functions $\rho_{x}e^{\beta\phi_{x}}$ in the
domains $\frac{\sigma}{2}<x<\frac{\sigma}{2}+\Delta$ and $\frac{5\sigma}%
{2}<x<\frac{5\sigma}{2}+\Delta$ are trivially related.

\subsubsection{Differential relations}

Multiplying Eq.(\ref{d1}) by $e^{\beta\phi_{x}}$ and taking the derivative
gives a new set of relations
\begin{subequations}
\begin{align}
\Lambda^{2}Z^{\left(  2\right)  }\left[  \phi\right]  \frac{d}{dx}\left(
e^{\beta\phi_{x}}\rho_{x}\right)   &  =-e^{-\beta\phi_{x+\sigma}%
},\;\;\;\;\;\;\;\;\;\;\frac{\sigma}{2}<x<\frac{3\sigma}{2}\\
\Lambda^{2}Z^{\left(  2\right)  }\left[  \phi\right]  \frac{d}{dx}\left(
e^{\beta\phi_{x}}\rho_{x}\right)   &  =e^{-\beta\phi_{x-\sigma}}-e^{-\beta
\phi_{x+\sigma}},\;\frac{3\sigma}{2}<x<\frac{3\sigma}{2}+\Delta\nonumber\\
\Lambda^{2}Z^{\left(  2\right)  }\left[  \phi\right]  \frac{d}{dx}\left(
e^{\beta\phi_{x}}\rho_{x}\right)   &  =e^{-\beta\phi_{x-\sigma}}%
,\;\;\;\;\;\;\;\;\;\;\;\frac{3\sigma}{2}+\Delta<x<\frac{5\sigma}{2}%
+\Delta\nonumber
\end{align}
and shifting the spatial variable in the first and third of these gives
\end{subequations}
\begin{subequations}
\label{e1}%
\begin{align}
\Lambda^{2}Z^{\left(  2\right)  }\left[  \phi\right]  \frac{d}{dx}\left(
e^{\beta\phi_{x-\sigma}}\rho_{x-\sigma}\right)   &  =-e^{-\beta\phi_{x}%
},\;\;\;\;\;\;\;\;\;\;\frac{3\sigma}{2}<x<\frac{5\sigma}{2}\label{ea}\\
\Lambda^{2}Z^{\left(  2\right)  }\left[  \phi\right]  \frac{d}{dx}\left(
e^{\beta\phi_{x}}\rho_{x}\right)   &  =e^{-\beta\phi_{x-\sigma}}-e^{-\beta
\phi_{x+\sigma}},\;\frac{3\sigma}{2}<x<\frac{3\sigma}{2}+\Delta\label{eb}\\
\Lambda^{2}Z^{\left(  2\right)  }\left[  \phi\right]  \frac{d}{dx}\left(
e^{\beta\phi_{x+\sigma}}\rho_{x+\sigma}\right)   &  =e^{-\beta\phi_{x}%
},\;\;\;\;\;\;\;\;\;\;\;\;\;\frac{\sigma}{2}+\Delta<x<\frac{3\sigma}{2}%
+\Delta. \label{ec}%
\end{align}
Taking advantage of overlaps between the regions in Eq.(\ref{d1}) and
Eq.(\ref{e1}) and repeatedly using the duality relation and shifts of the
spatial arguments (see Supplementary Text\cite{SI}) results in a closed
systems of equations which can be partially solved with the result:
\end{subequations}
\begin{subequations}
\label{g}%
\begin{align}
\frac{d}{dx}\frac{\rho_{x+\sigma}}{\frac{d}{dx}\left(  e^{\beta\phi_{x}}%
\rho_{x}\right)  }  &  =\frac{\rho_{x+2\sigma}}{A-\rho_{x}e^{\beta\phi_{x}}%
}-e^{-\beta\phi_{x}},\;\;\;\;\;\;\;\;\;\;\;\;\;\;\;\;\;\;\;\;\frac{\sigma}%
{2}<x<\frac{\sigma}{2}+\Delta\equiv D_{1}\label{g1}\\
e^{-\beta\phi_{x}}  &  =e^{-\beta\phi_{\frac{\sigma}{2}+\Delta}}\frac{\rho
_{x}}{\rho_{_{\frac{\sigma}{2}+\Delta}}}\exp\left(  \int_{\frac{\sigma}%
{2}+\Delta}^{x}\frac{\rho_{y+\sigma}}{\lambda_{y}\left[  \rho\right]
}dy\right)  ,\;\;\frac{\sigma}{2}+\Delta<x<\frac{3\sigma}{2}\equiv
D_{2}\label{g2}\\
e^{-\beta\phi_{x}}  &  =\pm\Lambda^{2}Z^{\left(  2\right)  }\left[
\phi\right]  \frac{d}{dx}\left(  e^{\beta\phi_{x\pm\sigma}}\rho_{x\pm\sigma
}\right)  ,\;\;\;\;\;\;\;\;\;\;\;\frac{3\sigma}{2}<x<\frac{3\sigma}{2}%
+\Delta\equiv D_{3}\label{g3}\\
e^{-\beta\phi_{x}}  &  =e^{-\beta\phi_{\frac{5\sigma}{2}}}\frac{\rho_{x}}%
{\rho_{\frac{5\sigma}{2}}}\exp\left(  \int_{x-\sigma}^{\frac{3\sigma}{2}}%
\frac{\rho_{y}}{\lambda_{y}\left[  \rho\right]  }dy\right)  ,\;\;\;\;\frac
{3\sigma}{2}+\Delta<x<\frac{5\sigma}{2}\equiv D_{4}\label{g4}\\
e^{-\beta\phi_{x}}  &  =\frac{\rho_{x}}{A-e^{\beta\phi_{x-2\sigma}}%
\rho_{x-2\sigma}},\;\;\;\;\;\;\;\;\;\;\;\;\;\;\;\;\;\;\;\;\;\;\frac{5\sigma
}{2}<x<\frac{5\sigma}{2}+\Delta\equiv D_{5} \label{g5}%
\end{align}
with%
\end{subequations}
\begin{align}
\lambda_{x}\left[  \rho\right]   &  =\lambda_{\frac{\sigma}{2}+\Delta}%
+\int_{\frac{\sigma}{2}+\Delta}^{\frac{3\sigma}{2}+\Delta}\rho_{y}dy-\int
_{x}^{x+\sigma}\rho_{y}dy=\lambda_{\frac{\sigma}{2}+\Delta}+\int_{\frac
{\sigma}{2}+\Delta}^{x}\rho_{y}dy-\int_{\frac{3\sigma}{2}+\Delta}^{x+\sigma
}\rho_{y}dy\\
\Lambda^{2}Z^{\left(  2\right)  }\left[  \phi\right]   &  =\frac{1}%
{e^{\beta\phi_{\frac{5\sigma}{2}}}\rho_{\frac{5\sigma}{2}}}\frac{1}%
{e^{\beta\phi_{\frac{\sigma}{2}+\Delta}}\rho_{_{\frac{\sigma}{2}+\Delta}}%
}\lambda_{\frac{\sigma}{2}+\Delta}\left[  \rho\right]  \exp\left(  \int
_{\frac{\sigma}{2}+\Delta}^{\frac{3\sigma}{2}}\frac{\rho_{y}}{\lambda
_{y}\left[  \rho\right]  }dy\right) \nonumber
\end{align}
Names have been assigned to various domains and the physical significance of
these divisions is as follows: the domain $D_{3}$ is the only one which both
hard rods can visit. When the rightmost rod is in this range, the leftmost is
confined to $D_{1}$ and when the leftmost is in the overlap range, the
rightmost is confined to $D_{5}$. The leftmost rod can be in the range $D_{2}$
when the when the rightmost rod is not in the overlap region, $D_{3}$, and
vice versa for $D_{4}$. In the course of solving the equations, the following
constraints are generated:%
\begin{subequations}
\begin{align}
\int_{\frac{\sigma}{2}}^{\frac{\sigma}{2}+\Delta}e^{-\beta\phi_{y}}dy  &
=e^{-\beta\phi_{\frac{\sigma}{2}+\Delta}}\frac{1}{\rho_{_{\frac{\sigma}%
{2}+\Delta}}}\lambda_{\frac{\sigma}{2}+\Delta}\label{xtra}\\
\int_{\frac{5\sigma}{2}}^{\frac{5\sigma}{2}+\Delta}e^{-\beta\phi_{y}}dy  &
=e^{-\beta\phi_{\frac{5\sigma}{2}}}\frac{1}{\rho_{\frac{5\sigma}{2}}}%
\lambda_{\frac{3\sigma}{2}}\nonumber\\
-\frac{\rho_{\frac{\sigma}{2}+\Delta}e^{\beta\phi_{\frac{\sigma}{2}+\Delta}}%
}{\left(  \rho_{x}e^{\beta\phi_{x}}\right)  _{\frac{\sigma}{2}+\Delta}%
^{\prime}}\rho_{\frac{3\sigma}{2}+\Delta}+\int_{\frac{\sigma}{2}+\Delta
}^{\frac{3\sigma}{2}}\rho_{y}dy  &  =\frac{\rho_{\frac{5\sigma}{2}}%
e^{\beta\phi_{\frac{5\sigma}{2}}}}{\left(  \rho_{x}e^{\beta\phi_{x}}\right)
_{\frac{5\sigma}{2}}^{\prime}}\rho_{\frac{3\sigma}{2}}+\int_{\frac{3\sigma}%
{2}+\Delta}^{\frac{5\sigma}{2}}\rho_{z}dz.\nonumber
\end{align}
The solution of Eq.(\ref{g}) involves 6 integration constants:\ two from the
ode in domain \ $D_{1}$, $e^{-\beta\phi_{\frac{\sigma}{2}+\Delta}}%
,e^{-\beta\phi_{\frac{5\sigma}{2}}},\lambda_{\frac{\sigma}{2}+\Delta}$ and
$A$. Continuity of the quantity $\rho_{x}e^{\beta\phi_{x}}$ across the
boundaries of the domains gives four conditions and the first two of
Eq.(\ref{xtra}) already give six. There is also the definition of $A$,
Eq(\ref{dual}), the jump conditions, Eq.(\ref{jump}), and the last of
Eq.(\ref{xtra}): clearly, many of these are redundant. In fact, based on the
solution given here, one can easily show that the jump conditions are
automatically satisfied and that the evaluation of $A$ from its definition
ends in a tautology giving no new information. Also, the second of the
relations in Eq.(\ref{xtra}) follows from the first as is easily shown using
Eq.(\ref{eb}) and the third relation also follows from the definitions. So, in
the end, there are only the four continuity relations and the first of
Eq.(\ref{xtra}) and the indeterminacy of the parameter $A$ represents the
gauge freedom of the potential.

As a simple illustration of these results, note first that in the case of no
field (or, more generally, a constant field), the density is a
piece-wise-linear function
\end{subequations}
\begin{subequations}
\begin{align}
\Lambda_{T}^{2}Z^{\left(  2\right)  }\left[  \phi\right]  \rho_{x}  &
=\left(  \Delta-x+\frac{3}{2}\sigma\right)
,\;\;\;\;\;\;\;\;\;\;\;\;\;\;\;\;\;\;\;\;\;\;\frac{\sigma}{2}<x<\frac{3\sigma
}{2}\\
&  =\Delta,\;\frac{3\sigma}{2}<x<\frac{3\sigma}{2}+\Delta\nonumber\\
&  =\left(  x-\frac{3}{2}\sigma\right)
,\;\;\;\;\;\;\;\;\;\;\;\;\;\;\;\;\;\;\;\;\;\;\;\;\frac{3\sigma}{2}%
+\Delta<x<\frac{5\sigma}{2}+\Delta.\nonumber
\end{align}
with%
\end{subequations}
\begin{equation}
\Lambda_{T}^{2}Z^{\left(  2\right)  }\left[  \phi\right]  =\frac{1}{2}\left(
\Delta+\sigma\right)  ^{2}%
\end{equation}
The field as a function of the density can be solved analytically for the case
of a constant density $\rho_{x}=\frac{2}{2\sigma+\Delta}$ throughout
$\frac{\sigma}{2}<x<\frac{5\sigma}{2}+\Delta$ with the result%
\begin{subequations}
\begin{align}
e^{-\beta\phi_{x}}  &  =\frac{\rho}{A}\left(  1+e^{B\left(  x-\frac{1}%
{2}\sigma-\frac{\Delta}{2}\right)  }\right)
,\;\;\;\;\;\;\;\;\;\;\;\;\;\;\;\;\;\;\;\;\;\;\;\;\frac{\sigma}{2}%
<x<\frac{\sigma}{2}+\Delta\\
&  =\frac{\rho}{A}\frac{B\Delta}{B\Delta-2}e^{-\frac{2}{\Delta}\left(
\frac{\sigma}{2}+\Delta-x\right)  }%
,\;\;\;\;\;\;\;\;\;\;\;\;\;\;\;\;\;\;\;\;\;\;\;\frac{\sigma}{2}+\Delta
<x<\frac{3\sigma}{2}\nonumber\\
&  =\frac{\rho}{A}\frac{\left(  B\Delta\right)  ^{3}}{4\left(  B\Delta
-2\right)  }e^{-\frac{2}{\Delta}\left(  \Delta-\sigma\right)  }\frac
{e^{B\left(  x-\frac{3}{2}\sigma\right)  }}{\left(  1+e^{B\left(  x-\frac
{3}{2}\sigma-\frac{\Delta}{2}\right)  }\right)  ^{2}},\;\frac{3\sigma}%
{2}<x<\frac{3\sigma}{2}+\Delta\nonumber\\
&  =\frac{\rho}{A}\frac{B\Delta}{B\Delta-2}e^{-\frac{2}{\Delta}\left(
x-\frac{5}{2}\sigma\right)  }%
,\;\;\;\;\;\;\;\;\;\;\;\;\;\;\;\;\;\;\;\;\;\;\;\;\frac{3\sigma}{2}%
+\Delta<x<\frac{5\sigma}{2}\nonumber\\
&  =\frac{\rho}{A}\left(  1+e^{B\left(  \frac{5}{2}\sigma+\frac{\Delta}%
{2}-x\right)  }\right)
,\;\;\;\;\;\;\;\;\;\;\;\;\;\;\;\;\;\;\;\;\;\;\;\;\;\frac{5\sigma}{2}%
<x<\frac{5\sigma}{2}+\Delta\nonumber
\end{align}
where, $A$ is the expected arbitrary constant (so that the family of equivalent potentials is $\phi_x -k_BT \ln A$), the constant $B$ is determined from $B\Delta=2+2W\left(  e^{-1}\right)
\approx2.\,\allowbreak556\,92$ where $W(x)$ is the Lambert W-function and the
partition function is
\end{subequations}
\begin{equation}
\Lambda^{2}Z^{\left(  2\right)  }\left[  \phi\right]  =\left(  \frac{B\Delta
}{A\left(  B\Delta-2\right)  }\right)  ^{2}\frac{1}{2}\rho\Delta e^{-\frac
{2}{\Delta}\left(  \Delta-\sigma\right)  }.
\end{equation}

\subsubsection{The Helmholtz functional}

The Helmholtz functional is determined from Eq.(\ref{f}) as in the case of the
ideal gas (see Supplementary Text\cite{SI}). Not all contributions can be
explicitly worked out but a useful result is still possible in the form
\begin{equation}
F\left[  \rho\right]  =\int_{\frac{\sigma}{2}}^{\frac{5\sigma}{2}}\rho
_{x}\left(  \ln\Lambda_{T}\rho_{x}-1\right)  dx-\int_{\sigma}^{2\sigma}%
s_{x}\left[  \rho\right]  \ln\left(  1-\eta_{x}\left[  \rho\right]  \right)
dx+O\left(  \frac{\Delta}{\sigma}\right)  . \label{result}%
\end{equation}
In fact the limit $\frac{\Delta}{\sigma}=0$ can be derived directly but there
are certain ambiguities which, in this extended calculation, resolve as terms
which separately diverge in the limit $\frac{\Delta}{\sigma}\rightarrow0$ but
which are collectively finite for all $\frac{\Delta}{\sigma}$. This was, in
fact, the reason for considering this more general system. It is interesting
to note that the origin of the singularities lies in the physical fact that
for a cavity of length $3\sigma$, the center of one rod is confined to
$\left[  \frac{\sigma}{2},\frac{3\sigma}{2}\right]  $ and that of the second
to $\left[  \frac{3\sigma}{2},\frac{5\sigma}{2}\right]  $ so that $\rho
_{\frac{3\sigma}{2}}=0$ and $\eta_{\sigma}\left[  \rho\right]  =\eta_{2\sigma
}\left[  \rho\right]  =1$ , the latter fact leading to difficulties with the
log in Eq.(\ref{result}).

\section{Conclusions}

It has been shown that the mapping between the external field and the density
in both the canonical and grand-canonical ensembles is virtually identical
with the only difference being an unimportant freedom in the canonical
ensemble to shift the field arbitrarily (and this freedom can be removed by
imposing a gauge condition). As a consequence, DFT in the two ensembles is
formally identical and this is explicitly seen in the case of the ideal gas
for which the functionals are almost the same in the two ensembles. Beyond the
ideal gas, there are only a few systems for which exact results have been
derived in the grand-canonical ensemble: hard particles in small cavities that
can only hold a single particle and, at the other extreme, hard rods in one
dimension with no constraint on the geometry (and slight generalizations, such
as sticky hard rods). It was shown here that for chains of small cavities,
results for small cavities can be easily obtained in the canonical ensemble.
These can do doubt be extended, as in the grand-canonical ensemble, to other
interesting topologies\cite{lutsko2020}. Finally, the important case of
hard-rods in one dimension was considered where the general exact result for
the grand-canonical ensemble is known. In this case, the canonical ensemble
seems to be more difficult to work with and in fact, the problem of even two
hard rods in a restricted geometry turns out to be difficult to solve
explicitly although a formal solution was constructed. A remarkable aspect of
the resulting Helmholtz functional was the close similarity it has to the
general grand-canonical result. This can be made even more apparent by writing
them together:%
\begin{align}
F\left[  \rho\right]   &  =\int_{\frac{\sigma}{2}}^{\frac{5\sigma}{2}}\rho
_{x}\left(  \ln\Lambda_{T}\rho_{x}-1\right)  dx-\int_{\sigma}^{2\sigma}%
s_{x}\left[  \rho\right]  \ln\left(  1-\eta_{x}\left[  \rho\right]  \right)
dx+O\left(  \frac{\Delta}{\sigma}\right)  ,\;\text{(C)}\\
F\left[  \rho\right]   &  =\int_{\frac{\sigma}{2}}^{\frac{5\sigma}{2}}\rho
_{x}\left(  \ln\Lambda_{T}\rho_{x}-1\right)  dx-\int_{0}^{3\sigma}s_{x}\left[
\rho\right]  \ln\left(  1-\eta_{x}\left[  \rho\right]  \right)
dx,\;\;\;\;\;\;\;\;\;\;\;\;\text{(GC)}\nonumber
\end{align}
so that one sees that to leading order in $\frac{\Delta}{\sigma}$, the only
difference is in the limits of the integrals. Nevertheless, the overall
complexity of the full result is far more involved than for the
grand-canonical ensemble, thus highlighting important differences between them.

Finally, one can only speculate on the broader implications of these results.
Clearly, cDFT exists equally rigorously in both the canonical and
grand-canonical ensembles. The most important difference between them is in
the variety of exact results available on which to base models and even then,
the gap is not as large as might be expected. Indeed, in the examples
considered here, the functionals show certain similarities of structure. While
this might have been anticipated for the ideal gas, it is very surprising that
even for a very small system such as the case of two hard-rods, the canonical
and grand-canonical Helmholtz functionals can be so similar. While this does
not rigorously justify using grand-canonical functionals in canonical models,
it does suggest that do so - for lack of better options - is not unreasonable.

\begin{acknowledgments}
I thank James Dufty, Sam Trickey and Bob Evans for useful comments and for pointing out relevant prior work. 
This work was supported by the European Space Agency (ESA) and the Belgian
Federal Science Policy Office (BELSPO) in the framework of the PRODEX
Programme, contract number ESA AO-2004-070.
\end{acknowledgments}

\bibliography{canonical}

\appendix

\section*{The Helmholtz functional}

For completeness, the expression for the Helmholtz functional for two rods in
a cavity in the canonical ensemble as derived in the Supplementary
Text\cite{SI} is quoted here:
\begin{align}
\beta F_{2}\left[  \rho\right]   &  =F_{2}^{\text{ideal}}\left[  \rho
_{N}\right]  -\int_{\sigma+\Delta}^{2\sigma}s_{x}[\rho]\ln(\Delta\lambda
+\eta_{x}[\rho])dx\\
&  +\left(  1-\int_{\frac{\sigma}{2}+\Delta}^{\frac{3\sigma}{2}}\rho
_{x}dx\right)  \ln\frac{e^{\beta\phi_{\frac{\sigma}{2}+\Delta}}\rho
_{_{\frac{\sigma}{2}+\Delta}}}{\lambda_{\frac{\sigma}{2}+\Delta}^{1/2}%
}+\left(  1-\int_{\frac{3\sigma}{2}+\Delta}^{\frac{5\sigma}{2}}\rho
_{x}dx\right)  \ln\frac{e^{\beta\phi_{\frac{5\sigma}{2}}}\rho_{\frac{5\sigma
}{2}}}{\lambda_{\frac{3\sigma}{2}}^{1/2}}\nonumber\\
&  +\frac{1}{2}\left(  \int_{\frac{\sigma}{2}+\Delta}^{\frac{3\sigma}{2}}%
\rho_{y}dy+\lambda_{\frac{\sigma}{2}+\Delta}-1\right)  \left(  \int
_{\frac{\sigma}{2}+\Delta}^{\frac{3\sigma}{2}}\frac{\rho_{y+\sigma}+\rho_{y}%
}{\lambda_{y}}dy\right)  -1\nonumber\\
&  -\sum_{a=0}^{2}\int_{\frac{\sigma}{2}+a\sigma}^{\frac{\sigma}{2}%
+a\sigma+\Delta}\rho_{x}\left(  \ln e^{\beta\phi_{x}}\rho_{x}-\frac{1}%
{2}\right)  dx\nonumber
\end{align}
where $F_{2}^{\text{ideal}}$ is the ideal gas functional,Eq.(\ref{idealgas}),
$\eta_{x}[\rho],s_{x}[\rho]$ are components of the Percus functional,
Eq.(\ref{Percus2}), and
\begin{equation}
\Delta\lambda\equiv\lambda_{\frac{\sigma}{2}+\Delta}+\int_{\frac{\sigma}%
{2}+\Delta}^{\frac{3\sigma}{2}+\Delta}\rho_{y}dy-1
\end{equation}
Finally, it is important to note that these expressions are not as explicit as
they appear since the quantities $\phi_{\frac{\sigma}{2}+\Delta},\phi
_{\frac{5\sigma}{2}},\lambda_{\frac{\sigma}{2}+\Delta}$ are all functionals of
the density as well as, of course, the explicit contributions in $\phi
_{x}[\rho]$.

\newpage
\section*{Supplementary Text}

\section{Canonical Cavity: the field in terms of the density}

The problem concerns two hard rods of length $\sigma$ in a cavity defined by
$0\leq x\leq L=3\sigma+\Delta$ with $\Delta<\sigma$ . This means that the
centers of a single hard rod is confined to the interval $\frac{\sigma}{2}\leq
x\leq\frac{5\sigma}{2}+\Delta$ and so is defined by an external field that is
infinite everywhere outside this range. For convenience, we recall the
important preliminary results from the main text. The density is
\begin{subequations}
\label{d1}%
\begin{align}
\Lambda_{T}^{2}Z^{\left(  2\right)  }\left[  \phi\right]  \rho_{x} &
=e^{-\beta\phi_{x}}\int_{x+\sigma}^{\frac{5\sigma}{2}+\Delta}e^{-\beta\phi
_{y}}dy,\;\;\;\;\;\;\;\;\;\;\;\;\;\;\;\;\;\;\;\;\;\;\frac{\sigma}{2}%
<x<\frac{3\sigma}{2}\label{da}\\
&  =e^{-\beta\phi_{x}}\int_{\frac{\sigma}{2}}^{x-\sigma}e^{-\beta\phi_{y}%
}dy+e^{-\beta\phi_{x}}\int_{x+\sigma}^{\frac{5\sigma}{2}+\Delta}e^{-\beta
\phi_{y}}dy,\;\frac{3\sigma}{2}<x<\frac{3\sigma}{2}+\Delta\label{db}\\
&  =e^{-\beta\phi_{x}}\int_{\frac{\sigma}{2}}^{x-\sigma}e^{-\beta\phi_{y}%
}dy,\;\;\;\;\;\;\;\;\;\;\;\;\;\;\;\;\;\;\;\;\;\;\;\;\frac{3\sigma}{2}%
+\Delta<x<\frac{5\sigma}{2}+\Delta.\label{dc}%
\end{align}
and from these one easily demonstrates the jump conditions%
\end{subequations}
\begin{align}
\lim_{\epsilon\rightarrow0}\left(  \frac{d}{dx}\rho_{x}e^{\beta\phi_{x}%
}\right)  _{\frac{3\sigma}{2}+\epsilon} &  =\lim_{\epsilon\rightarrow0}\left(
\frac{d}{dx}\rho_{x}e^{\beta\phi_{x}}\right)  _{\frac{3\sigma}{2}-\epsilon
}+\frac{1}{\Lambda_{T}^{2}Z^{\left(  2\right)  }\left[  \phi\right]
}e^{-\beta\phi_{\frac{\sigma}{2}}}\label{jump}\\
\lim_{\epsilon\rightarrow0}\left(  \frac{d}{dx}\rho_{x}e^{\beta\phi_{x}%
}\right)  _{\frac{3\sigma}{2}+\Delta+\epsilon} &  =\lim_{\epsilon\rightarrow
0}\left(  \frac{d}{dx}\rho_{x}e^{\beta\phi_{x}}\right)  _{\frac{3\sigma}%
{2}+\Delta-\epsilon}+\frac{1}{\Lambda_{T}^{2}Z^{\left(  2\right)  }\left[
\phi\right]  }e^{-\beta\phi_{\frac{5\sigma}{2}+\Delta}}\nonumber
\end{align}
The duality relation is
\begin{equation}
\rho_{x}e^{\beta\phi_{x}}+\rho_{x+2\sigma}e^{\beta\phi_{x+2\sigma}}=\frac
{1}{\Lambda_{T}^{2}Z^{\left(  2\right)  }\left[  \phi\right]  }\int
_{\frac{\sigma}{2}}^{\frac{5\sigma}{2}+\Delta}e^{-\beta\phi_{y}}dy\equiv
A,\;\;\;\;\;\frac{\sigma}{2}<x<\frac{\sigma}{2}+\Delta\label{dual}%
\end{equation}
and the differential relations derived in the main text are
\begin{subequations}
\label{e1}%
\begin{align}
\Lambda^{2}Z^{\left(  2\right)  }\left[  \phi\right]  \frac{d}{dx}\left(
e^{\beta\phi_{x-\sigma}}\rho_{x-\sigma}\right)   &  =-e^{-\beta\phi_{x}%
},\;\;\;\;\;\;\;\;\;\;\frac{3\sigma}{2}<x<\frac{5\sigma}{2}\label{ea}\\
\Lambda^{2}Z^{\left(  2\right)  }\left[  \phi\right]  \frac{d}{dx}\left(
e^{\beta\phi_{x}}\rho_{x}\right)   &  =e^{-\beta\phi_{x-\sigma}}-e^{-\beta
\phi_{x+\sigma}},\;\frac{3\sigma}{2}<x<\frac{3\sigma}{2}+\Delta\label{eb}\\
\Lambda^{2}Z^{\left(  2\right)  }\left[  \phi\right]  \frac{d}{dx}\left(
e^{\beta\phi_{x+\sigma}}\rho_{x+\sigma}\right)   &  =e^{-\beta\phi_{x}%
},\;\;\;\;\;\;\;\;\;\;\;\;\;\frac{\sigma}{2}+\Delta<x<\frac{3\sigma}{2}%
+\Delta.\label{ec}%
\end{align}

\subsection{Solving the canonical cavity\label{AppA}}

Using Eq.(\ref{ea}) one can rewrite Eq.(\ref{db}) as
\end{subequations}
\begin{subequations}
\begin{equation}
\rho_{x}=-\frac{d}{dx}\left(  e^{\beta\phi_{x-\sigma}}\rho_{x-\sigma}\right)
\left\{  \int_{\frac{\sigma}{2}}^{x-\sigma}e^{-\beta\phi_{y}}dy+\int
_{x+\sigma}^{\frac{5\sigma}{2}+\Delta}e^{-\beta\phi_{y}}dy\right\}
,\;\frac{3\sigma}{2}<x<\frac{3\sigma}{2}+\Delta
\end{equation}
So
\end{subequations}
\begin{subequations}
\begin{equation}
\frac{d}{dx}\frac{\rho_{x}}{\frac{d}{dx}\left(  e^{\beta\phi_{x-\sigma}}%
\rho_{x-\sigma}\right)  }=-e^{-\beta\phi_{x-\sigma}}+e^{-\beta\phi_{x+\sigma}%
},\;\frac{3\sigma}{2}<x<\frac{3\sigma}{2}+\Delta
\end{equation}
and shifting gives
\end{subequations}
\begin{subequations}
\begin{equation}
\frac{d}{dx}\frac{\rho_{x+\sigma}}{\frac{d}{dx}\left(  e^{\beta\phi_{x}}%
\rho_{x}\right)  }=-e^{-\beta\phi_{x}}+e^{-\beta\phi_{x+2\sigma}}%
,\;\frac{\sigma}{2}<x<\frac{\sigma}{2}+\Delta
\end{equation}
and the duality relation allows this to be written as
\end{subequations}
\begin{subequations}
\begin{equation}
\frac{d}{dx}\frac{\rho_{x+\sigma}}{\frac{d}{dx}\left(  e^{\beta\phi_{x}}%
\rho_{x}\right)  }=\frac{\rho_{x+2\sigma}}{A-e^{\beta\phi_{x}}\rho_{x}%
}-e^{-\beta\phi_{x}},\;\frac{\sigma}{2}<x<\frac{\sigma}{2}+\Delta.
\end{equation}
Using it again to replace $e^{\beta\phi_{x}}\rho_{x}$ by $A-\rho_{x+2\sigma
}e^{\beta\phi_{x+2\sigma}}$and shifting gives
\end{subequations}
\begin{subequations}
\begin{equation}
\frac{d}{dx}\frac{\rho_{x-\sigma}}{\frac{d}{dx}\left(  e^{\beta\phi_{x}}%
\rho_{x}\right)  }=\frac{\rho_{x-2\sigma}}{A-e^{\beta\phi_{x}}\rho_{x}%
}-e^{-\beta\phi_{x}},\;\frac{5\sigma}{2}<x<\frac{5\sigma}{2}+\Delta.
\end{equation}
Eq.(\ref{ea}) then immediately gives%
\end{subequations}
\begin{equation}
e^{-\beta\phi_{x}}=-\Lambda^{2}Z^{\left(  2\right)  }\left[  \phi\right]
\frac{d}{dx}\left(  e^{\beta\phi_{x-\sigma}}\rho_{x-\sigma}\right)
=\Lambda^{2}Z^{\left(  2\right)  }\left[  \phi\right]  \frac{d}{dx}\left(
e^{\beta\phi_{x+\sigma}}\rho_{x+\sigma}\right)  ,\;\;\;\;\;\;\;\;\;\;\frac
{3\sigma}{2}<x<\frac{3\sigma}{2}+\Delta
\end{equation}
with the second equality following from duality.

To get the remaining intervals, start with Eq.(\ref{dc}) and use Eq.(\ref{ea})
to write
\begin{subequations}
\begin{equation}
\rho_{x}=-\frac{d}{dx}\left(  e^{\beta\phi_{x-\sigma}}\rho_{x-\sigma}\right)
\int_{\frac{\sigma}{2}}^{x-\sigma}e^{-\beta\phi_{y}}dy,\;\;\frac{3\sigma}%
{2}+\Delta<x<\frac{5\sigma}{2}.
\end{equation}
Differentiate and shift to get
\end{subequations}
\begin{equation}
\frac{d}{dx}\frac{\rho_{x+\sigma}}{\frac{d}{dx}\left(  e^{\beta\phi_{x}}%
\rho_{x}\right)  }=-e^{-\beta\phi_{x}},\;\;\frac{\sigma}{2}+\Delta
<x<\frac{3\sigma}{2}.
\end{equation}
The final interval follows from the same process using Eq.(\ref{da}) and
(\ref{ec})%
\begin{equation}
\frac{d}{dx}\frac{\rho_{x-\sigma}}{\frac{d}{dx}\left(  e^{\beta\phi_{x}}%
\rho_{x}\right)  }=-e^{-\beta\phi_{x}},\;\;\;\;\;\;\;\frac{3\sigma}{2}%
+\Delta<x<\frac{5\sigma}{2}.
\end{equation}
We can then summarize as
\begin{align}
\frac{d}{dx}\frac{\rho_{x+\sigma}}{\frac{d}{dx}\left(  e^{\beta\phi_{x}}%
\rho_{x}\right)  } &  =\frac{\rho_{x+2\sigma}}{A-\rho_{x}e^{\beta\phi_{x}}%
}-e^{-\beta\phi_{x}},\;\;\;\;\;\;\;\;\;\frac{\sigma}{2}<x<\frac{\sigma}%
{2}+\Delta\label{summary}\\
\frac{d}{dx}\frac{\rho_{x+\sigma}}{\frac{d}{dx}\left(  e^{\beta\phi_{x}}%
\rho_{x}\right)  } &  =-e^{\beta\phi_{x}}%
,\;\;\;\;\;\;\;\;\;\;\;\;\;\;\;\;\;\;\;\;\;\;\;\;\;\frac{\sigma}{2}%
+\Delta<x<\frac{3\sigma}{2}\nonumber\\
e^{-\beta\phi_{x}} &  =\pm\Lambda^{2}Z^{\left(  2\right)  }\left[
\phi\right]  \frac{d}{dx}\left(  e^{\beta\phi_{x\pm\sigma}}\rho_{x\pm\sigma
}\right)  ,\;\frac{3\sigma}{2}<x<\frac{3\sigma}{2}+\Delta\nonumber\\
\frac{d}{dx}\frac{\rho_{x-\sigma}}{\frac{d}{dx}\left(  e^{\beta\phi_{x}}%
\rho_{x}\right)  } &  =-e^{-\beta\phi_{x}}%
,\;\;\;\;\;\;\;\;\;\;\;\;\;\;\;\;\;\;\;\;\;\;\;\frac{3\sigma}{2}%
+\Delta<x<\frac{5\sigma}{2}\nonumber\\
\frac{d}{dx}\frac{\rho_{x-\sigma}}{\frac{d}{dx}\left(  e^{\beta\phi_{x}}%
\rho_{x}\right)  } &  =\frac{\rho_{x-2\sigma}}{A-e^{\beta\phi_{x}}\rho_{x}%
}-e^{-\beta\phi_{x}},\;\;\;\;\;\;\;\;\;\frac{5\sigma}{2}<x<\frac{5\sigma}%
{2}+\Delta\nonumber
\end{align}

\subsection{Explicit solutions}

Define%
\begin{equation}
w_{x}=\left(  e^{\beta\phi_{x}}\rho_{x}\right)  ^{-1}%
\end{equation}
and after substituting into the first of Eq.(\ref{summary}) and simplifying one
gets%
\begin{equation}
\frac{d}{dx}\left(  -\frac{w_{x}\rho_{x+\sigma}}{w_{x}^{\prime}}\right)
=\rho_{x+\sigma}-\rho_{x},\;\;\frac{\sigma}{2}+\Delta<x<\frac{3\sigma}{2}%
\end{equation}
so%
\begin{equation}
\frac{w_{x}\rho_{x+\sigma}}{w_{x}^{\prime}}=\lambda_{-}\left(  x\right)
\equiv\left(  \frac{w_{x}\rho_{x+\sigma}}{w_{x}^{\prime}}\right)
_{\frac{\sigma}{2}+\Delta}-\int_{\frac{\sigma}{2}+\Delta}^{x}\left(
\rho_{y+\sigma}-\rho_{y}\right)  dy,\;\;\frac{\sigma}{2}+\Delta<x<\frac
{3\sigma}{2}.
\end{equation}
Solving gives%
\begin{equation}
\ln w_{x}=\ln w_{\frac{\sigma}{2}+\Delta}+\int_{\frac{\sigma}{2}+\Delta}%
^{x}\frac{\rho_{z+\sigma}}{\lambda_{z}^{\left(  -\right)  }}dz,\;\;\frac
{\sigma}{2}+\Delta<x<\frac{3\sigma}{2}%
\end{equation}
or%
\begin{equation}
e^{-\beta\phi_{x}}=e^{-\beta\phi_{\frac{\sigma}{2}+\Delta}}\frac{\rho_{x}%
}{\rho_{\frac{\sigma}{2}+\Delta}}\exp\left(  \int_{\frac{\sigma}{2}+\Delta
}^{x}\frac{\rho_{z+\sigma}}{\lambda_{z}^{\left(  -\right)  }}dz\right)
,\;\;\frac{\sigma}{2}+\Delta<x<\frac{3\sigma}{2}%
\end{equation}
so that $e^{-\beta\phi_{\frac{\sigma}{2}+\Delta}}$ and $\lambda_{-}\left(
\frac{\sigma}{2}+\Delta\right)  $ are the expected two integration constants.
Note that since $\frac{d}{dx}\lambda_{-}\left(  x\right)  =\rho_{x}%
-\rho_{x+\sigma}$ one has the very useful relations%
\begin{equation}
\frac{\rho_{x}}{\rho_{\frac{\sigma}{2}+\Delta}}\exp\left(  \int_{\frac{\sigma
}{2}+\Delta}^{x}\frac{\rho_{z+\sigma}}{\lambda_{z}^{\left(  -\right)  }%
}dz\right)  =\frac{\lambda_{\frac{\sigma}{2}+\Delta}^{\left(  -\right)  }%
}{\rho_{\frac{\sigma}{2}+\Delta}}\frac{\rho_{x}}{\lambda_{x}^{\left(
-\right)  }}\exp\left(  \int_{\frac{\sigma}{2}+\Delta}^{x}\frac{\rho_{z}%
}{\lambda_{z}^{\left(  -\right)  }}dz\right)  =\frac{\lambda_{\frac{\sigma}%
{2}+\Delta}^{\left(  -\right)  }}{\rho_{\frac{\sigma}{2}+\Delta}}\frac{d}%
{dx}\exp\left(  \int_{\frac{\sigma}{2}+\Delta}^{x}\frac{\rho_{z}}{\lambda
_{z}^{\left(  -\right)  }}dz\right)  .
\end{equation}
Similarly, one finds
\begin{equation}
e^{-\beta\phi_{x}}=e^{-\beta\phi_{\frac{5\sigma}{2}}}\frac{\rho_{x}}%
{\rho_{\frac{5\sigma}{2}}}\exp\left(  -\int_{x}^{\frac{5\sigma}{2}}\frac
{\rho_{y-\sigma}}{\lambda_{y}^{\left(  +\right)  }}dy\right)  ,\;\;\;\;\;\frac
{3\sigma}{2}+\Delta<x<\frac{5\sigma}{2}%
\end{equation}
with%
\begin{equation}
\lambda_{x}^{\left(  +\right)  }=-\frac{\rho_{\frac{5\sigma}{2}}e^{\beta
\phi_{\frac{5\sigma}{2}}}}{\left(  \rho_{x}e^{\beta\phi_{x}}\right)
_{\frac{5\sigma}{2}}^{\prime}}\rho_{\frac{3\sigma}{2}}-\int_{\frac{3\sigma}%
{2}}^{\frac{5\sigma}{2}}\rho_{z}dz+\int_{x-\sigma}^{x}\rho_{z}dz
\end{equation}
and again, since $\frac{d}{dx}\lambda_{x}^{\left(  +\right)  }=\rho_{x}%
-\rho_{x-\sigma}$,
\begin{equation}
\frac{\rho_{x}}{\rho_{\frac{5\sigma}{2}}}\exp\left(  -\int_{x}^{\frac{5\sigma
}{2}}\frac{\rho_{y-\sigma}}{\lambda_{y}^{\left(  +\right)  }}dy\right)
=\frac{\lambda_{\frac{5\sigma}{2}}^{\left(  +\right)  }}{\rho_{\frac{5\sigma
}{2}}}\frac{\rho_{x}}{\lambda_{x}^{\left(  +\right)  }}\exp\left(  -\int
_{x}^{\frac{5\sigma}{2}}\frac{\rho_{y}}{\lambda_{y}^{\left(  +\right)  }%
}dy\right)  =\frac{\lambda_{\frac{5\sigma}{2}}^{\left(  +\right)  }}%
{\rho_{\frac{5\sigma}{2}}}\frac{d}{dx}\exp\left(  -\int_{x}^{\frac{5\sigma}%
{2}}\frac{\rho_{y}}{\lambda_{y}^{\left(  +\right)  }}dy\right)  .\label{R2}%
\end{equation}

As a final step, we use both of these together to verify the original relation
between density and field, starting with Eq.(\ref{da})%
\begin{align}
\Lambda_{T}^{2}Z^{\left(  2\right)  }\left[  \phi\right]  \rho_{x}  &
=e^{-\beta\phi_{x}}\int_{x+\sigma}^{\frac{5\sigma}{2}+\Delta}e^{-\beta\phi
_{y}}dy,\;\;\;\;\frac{\sigma}{2}+\Delta<x<\frac{3\sigma}{2}\\
&  =e^{-\beta\phi_{x}}\left(  \int_{\frac{5\sigma}{2}}^{\frac{5\sigma}%
{2}+\Delta}e^{-\beta\phi_{y}}dy+\int_{x+\sigma}^{\frac{5\sigma}{2}}%
e^{-\beta\phi_{y}}dy\right) \nonumber\\
&  =e^{-\beta\phi_{x}}\left(  \int_{\frac{5\sigma}{2}}^{\frac{5\sigma}%
{2}+\Delta}e^{-\beta\phi_{y}}dy+e^{-\beta\phi_{\frac{5\sigma}{2}}}%
\frac{\lambda_{\frac{5\sigma}{2}}^{\left(  +\right)  }}{\rho_{\frac{5\sigma
}{2}}}-e^{-\beta\phi_{\frac{5\sigma}{2}}}\exp\left(  -\int_{x+\sigma}%
^{\frac{5\sigma}{2}}\frac{\rho_{y}}{\lambda_{y}^{\left(  +\right)  }%
}dy\right)  \right) \nonumber
\end{align}
where the third line follows from recognizing that the integrand is restricted
to the interval $\frac{3\sigma}{2}+\Delta<x<\frac{5\sigma}{2}$ and using the
last line of \ref{R2}. Substituting the solution for the field in $D_{2}$ and
simplifying gives%
\begin{align}
\Lambda_{T}^{2}Z^{\left(  2\right)  }\left[  \phi\right]   &  =e^{-\beta
\phi_{\frac{\sigma}{2}+\Delta}}\frac{1}{\rho_{\frac{\sigma}{2}+\Delta}}%
\exp\left(  \int_{\frac{\sigma}{2}+\Delta}^{x}\frac{\rho_{z+\sigma}}%
{\lambda_{z}^{\left(  -\right)  }}dz\right)  ,\;\;\;\;\frac{\sigma}{2}%
+\Delta<x<\frac{3\sigma}{2}\\
&  \times\left(  \int_{\frac{5\sigma}{2}}^{\frac{5\sigma}{2}+\Delta}%
e^{-\beta\phi_{y}}dy+e^{-\beta\phi_{\frac{5\sigma}{2}}}\frac{\lambda
_{\frac{5\sigma}{2}}^{\left(  +\right)  }}{\rho_{\frac{5\sigma}{2}}}%
-e^{-\beta\phi_{\frac{5\sigma}{2}}}\frac{\lambda_{\frac{5\sigma}{2}}^{\left(
+\right)  }}{\rho_{\frac{5\sigma}{2}}}\exp\left(  -\int_{x+\sigma}%
^{\frac{5\sigma}{2}}\frac{\rho_{y}}{\lambda_{z}^{\left(  +\right)  }%
}dy\right)  \right) \nonumber
\end{align}
The left hand side is independent of $x$ so the right hand side must be also.
Taking a derivative with respect to $x$ and simplifying gives%
\begin{equation}
0=\int_{\frac{5\sigma}{2}}^{\frac{5\sigma}{2}+\Delta}e^{-\beta\phi_{y}%
}dy+e^{-\beta\phi_{\frac{5\sigma}{2}}}\frac{\lambda_{\frac{5\sigma}{2}%
}^{\left(  +\right)  }}{\rho_{\frac{5\sigma}{2}}}-e^{-\beta\phi_{\frac
{5\sigma}{2}}}\left(  1+\frac{\lambda_{x}^{\left(  -\right)  }}{\lambda
_{x+\sigma}^{\left(  +\right)  }}\right)  \frac{\lambda_{\frac{5\sigma}{2}%
}^{\left(  +\right)  }}{\rho_{\frac{5\sigma}{2}}}\exp\left(  -\int_{x+\sigma
}^{\frac{5\sigma}{2}}\frac{\rho_{y}}{\lambda_{z}^{\left(  +\right)  }%
}dy\right)
\end{equation}
Another derivative and simplification gives%
\begin{equation}
0=\rho_{x}\frac{\lambda_{x+\sigma}^{\left(  +\right)  }+\lambda_{x}^{\left(
-\right)  }}{\lambda_{x+\sigma}^{\left(  +\right)  2}}%
\end{equation}
so one concludes that $\lambda_{x}^{\left(  +\right)  }=-\lambda_{x-\sigma
}^{\left(  -\right)  }$ which then, from the previous relation implies%
\begin{equation}
\int_{\frac{5\sigma}{2}}^{\frac{5\sigma}{2}+\Delta}e^{-\beta\phi_{y}%
}dy+e^{-\beta\phi_{\frac{5\sigma}{2}}}\frac{\lambda_{\frac{5\sigma}{2}%
}^{\left(  +\right)  }}{\rho_{\frac{5\sigma}{2}}}=0
\end{equation}
The original relation then becomes an expression for the partition function,%
\begin{align}
\Lambda_{T}^{2}Z^{\left(  2\right)  }\left[  \phi\right]   &  =e^{-\beta
\phi_{\frac{\sigma}{2}+\Delta}}\frac{1}{\rho_{\frac{\sigma}{2}+\Delta}}%
\exp\left(  \int_{\frac{\sigma}{2}+\Delta}^{x}\frac{\rho_{z+\sigma}}%
{\lambda_{z}^{\left(  -\right)  }}dz\right)  \left(  -e^{-\beta\phi
_{\frac{5\sigma}{2}}}\frac{\lambda_{\frac{5\sigma}{2}}^{\left(  +\right)  }%
}{\rho_{\frac{5\sigma}{2}}}\exp\left(  \int_{x+\sigma}^{\frac{5\sigma}{2}%
}\frac{\rho_{y}}{\lambda_{z-\sigma}^{\left(  -\right)  }}dy\right)  \right)
,\;\;\;\;\frac{\sigma}{2}+\Delta<x<\frac{3\sigma}{2}\\
&  =e^{-\beta\phi_{\frac{\sigma}{2}+\Delta}}e^{-\beta\phi_{\frac{5\sigma}{2}}%
}\frac{1}{\rho_{\frac{\sigma}{2}+\Delta}}\frac{\lambda_{\frac{3\sigma}{2}%
}^{\left(  -\right)  }}{\rho_{\frac{5\sigma}{2}}}\exp\left(  \int
_{\frac{\sigma}{2}+\Delta}^{\frac{3\sigma}{2}}\frac{\rho_{z+\sigma}}%
{\lambda_{z}^{\left(  -\right)  }}dz\right) \nonumber
\end{align}

One can perform the same exercise beginning with Eq.(\ref{dc}) with the only
new result being the condition%
\begin{equation}
\int_{\frac{\sigma}{2}}^{\frac{\sigma}{2}+\Delta}e^{-\beta\phi_{y}%
}dy=e^{-\beta\phi_{\frac{\sigma}{2}+\Delta}}\frac{\lambda_{\frac{\sigma}%
{2}+\Delta}^{\left(  -\right)  }}{\rho_{\frac{\sigma}{2}+\Delta}}%
\end{equation}

Thus, the result is that
\begin{align}
e^{-\beta\phi_{x}} &  =e^{-\beta\phi_{\frac{\sigma}{2}+\Delta}}\frac{\rho_{x}%
}{\rho_{\frac{\sigma}{2}+\Delta}}\exp\left(  \int_{\frac{\sigma}{2}+\Delta
}^{x}\frac{\rho_{z+\sigma}}{\lambda_{z}}dz\right)  ,\;\;\frac{\sigma}%
{2}+\Delta<x<\frac{3\sigma}{2}\label{ra}\\
e^{-\beta\phi_{x}} &  =e^{-\beta\phi_{\frac{5\sigma}{2}}}\frac{\rho_{x}}%
{\rho_{\frac{5\sigma}{2}}}\exp\left(  \int_{x}^{\frac{5\sigma}{2}}\frac
{\rho_{z-\sigma}}{\lambda_{z-\sigma}}dz\right)  ,\;\;\frac{3\sigma}{2}%
+\Delta<x<\frac{5\sigma}{2}\nonumber\\
\Lambda_{T}^{2}Z^{\left(  2\right)  }\left[  \phi\right]   &  =e^{-\beta
\phi_{\frac{\sigma}{2}+\Delta}}e^{-\beta\phi_{\frac{5\sigma}{2}}}\frac{1}%
{\rho_{\frac{\sigma}{2}+\Delta}}\frac{\lambda_{\frac{3\sigma}{2}}}{\rho
_{\frac{5\sigma}{2}}}\exp\left(  \int_{\frac{\sigma}{2}+\Delta}^{\frac
{3\sigma}{2}}\frac{\rho_{z+\sigma}}{\lambda_{z}}dz\right)  \nonumber\\
&  =e^{-\beta\phi_{\frac{\sigma}{2}+\Delta}}e^{-\beta\phi_{\frac{5\sigma}{2}}%
}\frac{1}{\rho_{\frac{\sigma}{2}+\Delta}}\frac{\lambda_{\frac{\sigma}%
{2}+\Delta}}{\rho_{\frac{5\sigma}{2}}}\exp\left(  \int_{\frac{\sigma}%
{2}+\Delta}^{\frac{3\sigma}{2}}\frac{\rho_{z}}{\lambda_{z}}dz\right)
\nonumber
\end{align}
with%
\begin{equation}
\lambda_{x}=\lambda_{\frac{\sigma}{2}+\Delta}+\int_{\frac{\sigma}{2}+\Delta
}^{\frac{3\sigma}{2}+\Delta}\rho_{y}dy-\int_{x}^{x+\sigma}\rho_{y}dy
\end{equation}
and the constraints
\begin{align}
\frac{e^{\beta\phi_{\frac{\sigma}{2}+\Delta}}\rho_{\frac{\sigma}{2}+\Delta}%
}{\left(  e^{\beta\phi_{x}}\rho_{x}\right)  _{\frac{\sigma}{2}+\Delta}%
^{\prime}}\rho_{\frac{3\sigma}{2}+\Delta}+\int_{\frac{\sigma}{2}+\Delta
}^{\frac{3\sigma}{2}+\Delta}\rho_{y}dy &  =\frac{\rho_{\frac{5\sigma}{2}%
}e^{\beta\phi_{\frac{5\sigma}{2}}}}{\left(  \rho_{x}e^{\beta\phi_{x}}\right)
_{\frac{5\sigma}{2}}^{\prime}}\rho_{\frac{3\sigma}{2}}+\int_{\frac{3\sigma}%
{2}}^{\frac{5\sigma}{2}}\rho_{z}dz\\
\lambda_{\frac{\sigma}{2}+\Delta} &  =\frac{e^{\beta\phi_{\frac{\sigma}%
{2}+\Delta}}\rho_{\frac{\sigma}{2}+\Delta}}{\left(  e^{\beta\phi_{x}}\rho
_{x}\right)  _{\frac{\sigma}{2}+\Delta}^{\prime}}\rho_{\frac{3\sigma}%
{2}+\Delta}\nonumber\\
\int_{\frac{5\sigma}{2}}^{\frac{5\sigma}{2}+\Delta}e^{-\beta\phi_{y}}dy &
=e^{-\beta\phi_{\frac{5\sigma}{2}}}\frac{\lambda_{\frac{3\sigma}{2}}}%
{\rho_{\frac{5\sigma}{2}}}\nonumber\\
\int_{\frac{\sigma}{2}}^{\frac{\sigma}{2}+\Delta}e^{-\beta\phi_{y}}dy &
=e^{-\beta\phi_{\frac{\sigma}{2}+\Delta}}\frac{\lambda_{\frac{\sigma}%
{2}+\Delta}}{\rho_{\frac{\sigma}{2}+\Delta}}\nonumber
\end{align}
The jump conditions are automatically satisfied.

\subsection{The definition of  the constant $A$ provides no new
information\label{AppC}}

We would like to simplify%
\begin{equation}
A=\frac{1}{\Lambda_{T}^{2}Z^{\left(  2\right)  }\left[  \phi\right]  }%
\int_{\frac{\sigma}{2}}^{\frac{5\sigma}{2}+\Delta}e^{-\beta\phi_{x}}dx
\end{equation}
so we consider the contribution from each domain separately. First,%
\begin{equation}
\Lambda_{T}^{2}Z^{\left(  2\right)  }\left[  \phi\right]  A_{1}=\int
_{\frac{\sigma}{2}}^{\frac{\sigma}{2}+\Delta}e^{-\beta\phi_{x}}dx=e^{-\beta
\phi_{\frac{\sigma}{2}+\Delta}}\frac{\lambda_{\frac{\sigma}{2}+\Delta}}%
{\rho_{\frac{\sigma}{2}+\Delta}}%
\end{equation}
but also%
\begin{align}
\Lambda_{T}^{2}Z^{\left(  2\right)  }\left[  \phi\right]  A_{1} &
=\int_{\frac{\sigma}{2}}^{\frac{\sigma}{2}+\Delta}e^{\beta\phi_{x+2\sigma}%
}dx-\int_{\frac{\sigma}{2}}^{\frac{\sigma}{2}+\Delta}\frac{d}{dx}\frac
{\rho_{x+\sigma}}{\frac{d}{dx}\left(  e^{\beta\phi_{x}}\rho_{x}\right)  }dx\\
&  =A_{5}-\frac{\rho_{\frac{3\sigma}{2}+\Delta}}{\frac{d}{dx}\left(
e^{\beta\phi_{x}}\rho_{x}\right)  _{\frac{\sigma}{2}+\Delta}}+\frac
{\rho_{\frac{3\sigma}{2}}}{\frac{d}{dx}\left(  e^{\beta\phi_{x}}\rho
_{x}\right)  _{\frac{\sigma}{2}}}\nonumber
\end{align}
Next%
\begin{align}
\Lambda_{T}^{2}Z^{\left(  2\right)  }\left[  \phi\right]  A_{2} &
=\int_{\frac{\sigma}{2}+\Delta}^{\frac{3\sigma}{2}}e^{-\beta\phi_{x}}dx\\
&  =e^{-\beta\phi_{\frac{\sigma}{2}+\Delta}}\frac{\lambda_{-}\left(
\frac{\sigma}{2}+\Delta\right)  }{\rho_{\frac{\sigma}{2}+\Delta}}\int
_{\frac{\sigma}{2}+\Delta}^{\frac{3\sigma}{2}}\frac{d}{dx}\exp\left(
\int_{\frac{\sigma}{2}+\Delta}^{x}\frac{\rho_{z}}{\lambda_{-}\left(  z\right)
}dz\right)  dx\nonumber\\
&  =e^{-\beta\phi_{\frac{\sigma}{2}+\Delta}}\frac{\lambda_{\frac{\sigma}%
{2}+\Delta}}{\rho_{\frac{\sigma}{2}+\Delta}}\left(  \exp\left(  \int
_{\frac{\sigma}{2}+\Delta}^{\frac{3\sigma}{2}}\frac{\rho_{z}}{\lambda_{z}%
}dz\right)  -1\right)  \nonumber
\end{align}
The third domain is easy%
\begin{align}
\Lambda_{T}^{2}Z^{\left(  2\right)  }\left[  \phi\right]  A_{3} &
=\int_{\frac{3\sigma}{2}}^{\frac{3\sigma}{2}+\Delta}e^{-\beta\phi_{x}}dx\\
&  =\Lambda^{2}Z^{\left(  2\right)  }\left[  \phi\right]  \left(  e^{\beta
\phi_{\frac{5\sigma}{2}+\Delta}}\rho_{\frac{5\sigma}{2}+\Delta}-e^{\beta
\phi_{\frac{5\sigma}{2}}}\rho_{\frac{5\sigma}{2}}\right)  \nonumber\\
&  =-\Lambda^{2}Z^{\left(  2\right)  }\left[  \phi\right]  \left(
e^{\beta\phi_{\frac{\sigma}{2}+\Delta}}\rho_{\frac{\sigma}{2}+\Delta}%
-e^{\beta\phi_{\frac{\sigma}{2}}}\rho_{\frac{\sigma}{2}}\right)  \nonumber
\end{align}
and the fourth gives%
\begin{align}
\Lambda_{T}^{2}Z^{\left(  2\right)  }\left[  \phi\right]  A_{4} &
=\int_{\frac{3\sigma}{2}+\Delta}^{\frac{5\sigma}{2}}e^{-\beta\phi_{x}%
}dx=e^{-\beta\phi_{\frac{5\sigma}{2}}}\frac{\lambda\left(  \frac{3\sigma}%
{2}\right)  }{\rho_{\frac{5\sigma}{2}}}\left(  \exp\left(  \int_{\frac
{3\sigma}{2}+\Delta}^{\frac{5\sigma}{2}}\frac{\rho_{y}}{\lambda\left(
y-\sigma\right)  }dy\right)  -1\right)  \\
&  =e^{-\beta\phi_{\frac{5\sigma}{2}}}\frac{\lambda_{\frac{\sigma}{2}+\Delta}%
}{\rho_{\frac{5\sigma}{2}}}\exp\left(  \int_{\frac{\sigma}{2}+\Delta}%
^{\frac{3\sigma}{2}}\frac{\rho_{y}}{\lambda_{y}}dy\right)  -e^{-\beta
\phi_{\frac{5\sigma}{2}}}\frac{\lambda\left(  \frac{3\sigma}{2}\right)  }%
{\rho_{\frac{5\sigma}{2}}}\nonumber
\end{align}
and the final one is
\begin{equation}
\Lambda_{T}^{2}Z^{\left(  2\right)  }\left[  \phi\right]  A_{5}=\int
_{\frac{5\sigma}{2}}^{\frac{5\sigma}{2}+\Delta}e^{-\beta\phi_{x}}%
dx=e^{-\beta\phi_{\frac{5\sigma}{2}}}\frac{\lambda\left(  \frac{3\sigma}%
{2}\right)  }{\rho_{\frac{5\sigma}{2}}}%
\end{equation}
So summing gives%
\begin{align}
\Lambda_{T}^{2}Z^{\left(  2\right)  }\left[  \phi\right]  A &  =e^{-\beta
\phi_{\frac{\sigma}{2}+\Delta}}\frac{\lambda_{\frac{\sigma}{2}+\Delta}}%
{\rho_{\frac{\sigma}{2}+\Delta}}\exp\left(  \int_{\frac{\sigma}{2}+\Delta
}^{\frac{3\sigma}{2}}\frac{\rho_{z}}{\lambda_{z}}dz\right)  +e^{-\beta
\phi_{\frac{5\sigma}{2}}}\frac{\lambda_{\frac{\sigma}{2}+\Delta}}{\rho
_{\frac{5\sigma}{2}}}\exp\left(  \int_{\frac{\sigma}{2}+\Delta}^{\frac
{3\sigma}{2}}\frac{\rho_{y}}{\lambda_{y}}dy\right)  \\
&  -\Lambda^{2}Z^{\left(  2\right)  }\left[  \phi\right]  \left(  e^{\beta
\phi_{\frac{\sigma}{2}+\Delta}}\rho_{\frac{\sigma}{2}+\Delta}-e^{\beta
\phi_{\frac{\sigma}{2}}}\rho_{\frac{\sigma}{2}}\right)  \nonumber
\end{align}
or%
\begin{equation}
\Lambda_{T}^{2}Z^{\left(  2\right)  }\left[  \phi\right]  A=\frac
{\lambda_{\frac{\sigma}{2}+\Delta}}{e^{\beta\phi_{\frac{\sigma}{2}+\Delta}%
}\rho_{\frac{\sigma}{2}+\Delta}\rho_{\frac{5\sigma}{2}}e^{\beta\phi
_{\frac{5\sigma}{2}}}}\exp\left(  \int_{\frac{\sigma}{2}+\Delta}%
^{\frac{3\sigma}{2}}\frac{\rho_{z}}{\lambda_{z}}dz\right)  \left(  \rho
_{\frac{5\sigma}{2}}e^{\beta\phi_{\frac{5\sigma}{2}}}+\rho_{\frac{\sigma}{2}%
}e^{\beta\phi_{\frac{\sigma}{2}}}\right)
\end{equation}
which are equal via the duality relation and using the known expression for
the partition function. Thus, this relation gives no new information.

\subsection{Proof that the partial integrals of the potential are not
independent}

Two constraints were derived above,
\begin{align}
I_{1} &  \equiv\int_{\frac{\sigma}{2}}^{\frac{\sigma}{2}+\Delta}e^{-\beta
\phi_{y}}dy=e^{-\beta\phi_{\frac{\sigma}{2}+\Delta}}\frac{\lambda
_{\frac{\sigma}{2}+\Delta}}{\rho_{\frac{\sigma}{2}+\Delta}}\\
I_{2} &  \equiv\int_{\frac{5\sigma}{2}}^{\frac{5\sigma}{2}+\Delta}%
e^{-\beta\phi_{y}}dy=e^{-\beta\phi_{\frac{5\sigma}{2}}}\frac{\lambda
_{\frac{3\sigma}{2}}}{\rho_{\frac{5\sigma}{2}}}\nonumber
\end{align}
and here the goal is to show that they are not independent. To begin,
integrate Rq.(\ref{eb}) over its range of validity to get%
\begin{equation}
\int_{\frac{3\sigma}{2}}^{\frac{3\sigma}{2}+\Delta}\left(  e^{-\beta
\phi_{x-\sigma}}-e^{-\beta\phi_{x+\sigma}}\right)  dx=\int_{\frac{3\sigma}{2}%
}^{\frac{3\sigma}{2}+\Delta}\Lambda^{2}Z^{\left(  2\right)  }\left[
\phi\right]  \frac{d}{dx}\left(  e^{\beta\phi_{x}}\rho_{x}\right)  dx
\end{equation}
or%
\begin{equation}
I_{1}-I_{2}=\Lambda^{2}Z^{\left(  2\right)  }\left[  \phi\right]  \left(
e^{\beta\phi_{\frac{3\sigma}{2}+\Delta}}\rho_{\frac{3\sigma}{2}+\Delta
}-e^{\beta\phi_{\frac{3\sigma}{2}}}\rho_{\frac{3\sigma}{2}}\right)
\end{equation}
We use the continuity of $e^{\beta\phi_{x}}\rho_{x}$ and the results given in
Eq.(\ref{ra}) for both the fields and the partition function to get%
\begin{equation}
I_{1}-I_{2}=e^{-\beta\phi_{\frac{\sigma}{2}+\Delta}}\frac{\lambda
_{\frac{\sigma}{2}+\Delta}}{\rho_{\frac{\sigma}{2}+\Delta}}-e^{-\beta
\phi_{\frac{5\sigma}{2}}}\frac{\lambda_{\frac{3\sigma}{2}}}{\rho
_{\frac{5\sigma}{2}}}%
\end{equation}
so that once one of the relations  is satisfied the other follows
automatically. 

\section{The Helmholtz functional}

As always, the strategy is to use the explicit expression for the field in
terms of the density to evaluate
\begin{equation}
\beta F_{N}\left[  \rho\right]  \equiv\beta A_{N}\left[  \phi_{N}\left[
\rho\right]  \right]  -\int\rho_{\mathbf{r}}\beta\phi_{N\mathbf{r}}\left[
\rho\right]  d\mathbf{r}.\;
\end{equation}
First, the free energy is
\begin{align}
\beta A_{2}\left[  \phi\left[  \rho\right]  \right]   &  =-\ln Z^{\left(
2\right)  }\left[  \phi\right] \\
&  =\ln e^{\beta\phi_{\frac{5\sigma}{2}}}\rho_{\frac{5\sigma}{2}}+\ln
e^{\beta\phi_{\frac{\sigma}{2}+\Delta}}\rho_{_{\frac{\sigma}{2}+\Delta}}%
-\ln\lambda_{\frac{\sigma}{2}+\Delta}-\int_{\frac{\sigma}{2}+\Delta}%
^{\frac{3\sigma}{2}}\frac{\rho_{y}}{\lambda_{y}}dy\nonumber
\end{align}
We will also need
\begin{align}
\int_{\frac{\sigma}{2}}^{\frac{5\sigma}{2}+\Delta}\rho_{x}\beta\phi_{x}dx  &
=\int_{\frac{\sigma}{2}}^{\frac{\sigma}{2}+\Delta}\rho_{x}\beta\phi_{x}%
dx+\int_{\frac{\sigma}{2}+\Delta}^{\frac{3\sigma}{2}}\rho_{x}\beta\phi_{x}dx\\
&  +\int_{\frac{3\sigma}{2}}^{\frac{3\sigma}{2}+\Delta}\rho_{x}\beta\phi
_{x}dx+\int_{\frac{3\sigma}{2}+\Delta}^{\frac{5\sigma}{2}}\rho_{x}\beta
\phi_{x}dx\nonumber\\
&  +\int_{\frac{5\sigma}{2}}^{\frac{5\sigma}{2}+\Delta}\rho_{x}\beta\phi
_{x}dx\nonumber
\end{align}
or, regrouping,%
\begin{align}
\int_{\frac{\sigma}{2}}^{\frac{5\sigma}{2}+\Delta}\rho_{x}\beta\phi_{x}dx  &
=\int_{\frac{\sigma}{2}+\Delta}^{\frac{3\sigma}{2}}\rho_{x}\beta\phi
_{x}dx+\int_{\frac{3\sigma}{2}+\Delta}^{\frac{5\sigma}{2}}\rho_{x}\beta
\phi_{x}dx\\
&  +\int_{\frac{\sigma}{2}}^{\frac{\sigma}{2}+\Delta}\rho_{x}\beta\phi
_{x}dx+\int_{\frac{3\sigma}{2}}^{\frac{3\sigma}{2}+\Delta}\rho_{x}\beta
\phi_{x}dx+\int_{\frac{5\sigma}{2}}^{\frac{5\sigma}{2}+\Delta}\rho_{x}%
\beta\phi_{x}dx\nonumber
\end{align}
So the Helmholtz functional is%
\begin{align}
\beta F\left[  \rho\right]   &  =\ln e^{\beta\phi_{\frac{5\sigma}{2}}}%
\rho_{\frac{5\sigma}{2}}+\ln e^{\beta\phi_{\frac{\sigma}{2}+\Delta}}%
\rho_{_{\frac{\sigma}{2}+\Delta}}-\ln\lambda_{\frac{\sigma}{2}+\Delta}%
-\int_{\frac{\sigma}{2}+\Delta}^{\frac{3\sigma}{2}}\frac{\rho_{y}}{\lambda
_{y}}dy\\
&  -\int_{\frac{\sigma}{2}+\Delta}^{\frac{3\sigma}{2}}\rho_{x}\beta\phi
_{x}dx-\int_{\frac{3\sigma}{2}+\Delta}^{\frac{5\sigma}{2}}\rho_{x}\beta
\phi_{x}dx\nonumber\\
&  -\int_{\frac{\sigma}{2}}^{\frac{\sigma}{2}+\Delta}\rho_{x}\beta\phi
_{x}dx-\int_{\frac{3\sigma}{2}}^{\frac{3\sigma}{2}+\Delta}\rho_{x}\beta
\phi_{x}dx-\int_{\frac{5\sigma}{2}}^{\frac{5\sigma}{2}+\Delta}\rho_{x}%
\beta\phi_{x}dx\nonumber
\end{align}

Substituting for the third line and simplifying gives%
\begin{align}
\beta F\left[  \rho\right]   &  =\left(  1-\int_{\frac{\sigma}{2}+\Delta
}^{\frac{3\sigma}{2}}\rho_{x}dx\right)  \ln e^{\beta\phi_{\frac{\sigma}%
{2}+\Delta}}\rho_{_{\frac{\sigma}{2}+\Delta}}+\left(  1-\int_{\frac{3\sigma
}{2}+\Delta}^{\frac{5\sigma}{2}}\rho_{x}dx\right)  \ln e^{\beta\phi
_{\frac{5\sigma}{2}}}\rho_{\frac{5\sigma}{2}}-\ln\lambda_{\frac{\sigma}%
{2}+\Delta}-\int_{\frac{\sigma}{2}+\Delta}^{\frac{3\sigma}{2}}\frac{\rho_{y}%
}{\lambda_{y}}dy\\
&  +\int_{\frac{\sigma}{2}+\Delta}^{\frac{3\sigma}{2}}\rho_{x}\ln\rho
_{x}dx+\int_{\frac{3\sigma}{2}+\Delta}^{\frac{5\sigma}{2}}\rho_{x}\ln\rho
_{x}dx\nonumber\\
&  +\int_{\frac{\sigma}{2}+\Delta}^{\frac{3\sigma}{2}}\rho_{x}\left(
\int_{\frac{\sigma}{2}+\Delta}^{x}\frac{\rho_{y+\sigma}}{\lambda_{y}%
}dy\right)  dx+\int_{\frac{3\sigma}{2}+\Delta}^{\frac{5\sigma}{2}}\rho
_{x}\left(  \int_{x-\sigma}^{\frac{3\sigma}{2}}\frac{\rho_{y}}{\lambda_{y}%
}dy\right)  dx\nonumber\\
&  -\int_{\frac{\sigma}{2}}^{\frac{\sigma}{2}+\Delta}\rho_{x}\beta\phi
_{x}dx-\int_{\frac{3\sigma}{2}}^{\frac{3\sigma}{2}+\Delta}\rho_{x}\beta
\phi_{x}dx-\int_{\frac{5\sigma}{2}}^{\frac{5\sigma}{2}+\Delta}\rho_{x}%
\beta\phi_{x}dx\nonumber
\end{align}
or, more succinctly,%
\begin{align}
\beta F\left[  \rho\right]   &  =\left(  1-\int_{\frac{\sigma}{2}+\Delta
}^{\frac{3\sigma}{2}}\rho_{x}dx\right)  \ln e^{\beta\phi_{\frac{\sigma}%
{2}+\Delta}}\rho_{_{\frac{\sigma}{2}+\Delta}}+\left(  1-\int_{\frac{3\sigma
}{2}+\Delta}^{\frac{5\sigma}{2}}\rho_{x}dx\right)  \ln e^{\beta\phi
_{\frac{5\sigma}{2}}}\rho_{\frac{5\sigma}{2}}-\ln\lambda_{\frac{\sigma}%
{2}+\Delta}-\int_{\frac{\sigma}{2}+\Delta}^{\frac{3\sigma}{2}}\frac{\rho_{y}%
}{\lambda_{y}}dy\\
&  +\int_{\frac{\sigma}{2}}^{\frac{5\sigma}{2}+\Delta}\rho_{x}\ln\rho
_{x}dx\nonumber\\
&  +\int_{\frac{\sigma}{2}+\Delta}^{\frac{3\sigma}{2}}\rho_{x}\left(
\int_{\frac{\sigma}{2}+\Delta}^{x}\frac{\rho_{y+\sigma}}{\lambda_{y}%
}dy\right)  dx+\int_{\frac{3\sigma}{2}+\Delta}^{\frac{5\sigma}{2}}\rho
_{x}\left(  \int_{x-\sigma}^{\frac{3\sigma}{2}}\frac{\rho_{y}}{\lambda_{y}%
}dy\right)  dx\nonumber\\
&  -\sum_{a=0}^{2}\int_{\frac{\sigma}{2}+a\sigma}^{\frac{\sigma}{2}%
+a\sigma+\Delta}\rho_{x}\left(  \beta\phi_{x}-\ln\rho_{x}\right)  dx\nonumber
\end{align}
We begin to analyze the third line by splitting it into two pieces%
\begin{align}
&  \int_{\frac{\sigma}{2}+\Delta}^{\frac{3\sigma}{2}}\rho_{x}\left(
\int_{\frac{\sigma}{2}+\Delta}^{x}\frac{\rho_{y+\sigma}}{\lambda_{y}%
}dy\right)  dx+\int_{\frac{3\sigma}{2}+\Delta}^{\frac{5\sigma}{2}}\rho
_{x}\left(  \int_{x-\sigma}^{\frac{3\sigma}{2}}\frac{\rho_{y}}{\lambda_{y}%
}dy\right)  dx\\
&  =\frac{1}{2}\int_{\frac{\sigma}{2}+\Delta}^{\frac{3\sigma}{2}}\rho
_{x}\left(  \int_{\frac{\sigma}{2}+\Delta}^{x}\frac{\rho_{y+\sigma}+\rho_{y}%
}{\lambda_{y}}dy\right)  dx+\frac{1}{2}\int_{\frac{3\sigma}{2}+\Delta}%
^{\frac{5\sigma}{2}}\rho_{x}\left(  \int_{x-\sigma}^{\frac{3\sigma}{2}}%
\frac{\rho_{y}+\rho_{y+\sigma}}{\lambda_{y}}dy\right)  dx\nonumber\\
&  +\frac{1}{2}\int_{\frac{\sigma}{2}+\Delta}^{\frac{3\sigma}{2}}\rho
_{x}\left(  \int_{\frac{\sigma}{2}+\Delta}^{x}\frac{\rho_{y+\sigma}-\rho_{y}%
}{\lambda_{y}}dy\right)  dx+\frac{1}{2}\int_{\frac{3\sigma}{2}+\Delta}%
^{\frac{5\sigma}{2}}\rho_{x}\left(  \int_{x-\sigma}^{\frac{3\sigma}{2}}%
\frac{\rho_{y}-\rho_{y+\sigma}}{\lambda_{y}}dy\right)  dx\nonumber
\end{align}
Recall that
\begin{equation}
\lambda_{x}=\lambda_{\frac{\sigma}{2}+\Delta}+\int_{\frac{\sigma}{2}+\Delta
}^{\frac{3\sigma}{2}+\Delta}\rho_{y}dy-\int_{x}^{x+\sigma}\rho_{y}dy
\end{equation}
so%
\begin{align}
&  \int_{\frac{\sigma}{2}+\Delta}^{\frac{3\sigma}{2}}\rho_{x}\left(
\int_{\frac{\sigma}{2}+\Delta}^{x}\frac{\rho_{y+\sigma}}{\lambda_{y}%
}dy\right)  dx+\int_{\frac{3\sigma}{2}+\Delta}^{\frac{5\sigma}{2}}\rho
_{x}\left(  \int_{x-\sigma}^{\frac{3\sigma}{2}}\frac{\rho_{y}}{\lambda_{y}%
}dy\right)  dx\\
&  =\frac{1}{2}\int_{\frac{\sigma}{2}+\Delta}^{\frac{3\sigma}{2}}\rho
_{x}\left(  \int_{\frac{\sigma}{2}+\Delta}^{x}\frac{\rho_{y+\sigma}+\rho_{y}%
}{\lambda_{y}}dy\right)  dx+\frac{1}{2}\int_{\frac{3\sigma}{2}+\Delta}%
^{\frac{5\sigma}{2}}\rho_{x}\left(  \int_{x-\sigma}^{\frac{3\sigma}{2}}%
\frac{\rho_{y}+\rho_{y+\sigma}}{\lambda_{y}}dy\right)  dx\nonumber\\
&  +\frac{1}{2}\int_{\frac{\sigma}{2}+\Delta}^{\frac{3\sigma}{2}}\rho_{x}%
\ln\frac{\lambda_{\frac{\sigma}{2}+\Delta}}{\lambda_{x}}dx+\frac{1}{2}%
\int_{\frac{3\sigma}{2}+\Delta}^{\frac{5\sigma}{2}}\rho_{x}\ln\frac
{\lambda_{\frac{3\sigma}{2}}}{\lambda_{x-\sigma}}dx\nonumber
\end{align}
So%
\begin{align}
\beta F\left[  \rho\right]   &  =\left(  1-\int_{\frac{\sigma}{2}+\Delta
}^{\frac{3\sigma}{2}}\rho_{x}dx\right)  \ln e^{\beta\phi_{\frac{\sigma}%
{2}+\Delta}}\rho_{_{\frac{\sigma}{2}+\Delta}}+\left(  1-\int_{\frac{3\sigma
}{2}+\Delta}^{\frac{5\sigma}{2}}\rho_{x}dx\right)  \ln e^{\beta\phi
_{\frac{5\sigma}{2}}}\rho_{\frac{5\sigma}{2}}-\ln\lambda_{\frac{\sigma}%
{2}+\Delta}-\int_{\frac{\sigma}{2}+\Delta}^{\frac{3\sigma}{2}}\frac{\rho_{y}%
}{\lambda_{y}}dy\\
&  +\int_{\frac{\sigma}{2}}^{\frac{5\sigma}{2}+\Delta}\rho_{x}\ln\rho
_{x}dx+\frac{1}{2}\int_{\frac{\sigma}{2}+\Delta}^{\frac{3\sigma}{2}}\rho
_{x}\ln\frac{\lambda_{\frac{\sigma}{2}+\Delta}}{\lambda_{x}}dx+\frac{1}{2}%
\int_{\frac{3\sigma}{2}+\Delta}^{\frac{5\sigma}{2}}\rho_{x}\ln\frac
{\lambda_{\frac{3\sigma}{2}}}{\lambda_{x-\sigma}}dx\nonumber\\
&  +\frac{1}{2}\int_{\frac{\sigma}{2}+\Delta}^{\frac{3\sigma}{2}}\rho
_{x}\left(  \int_{\frac{\sigma}{2}+\Delta}^{x}\frac{\rho_{y+\sigma}+\rho_{y}%
}{\lambda_{y}}dy\right)  dx+\frac{1}{2}\int_{\frac{3\sigma}{2}+\Delta}%
^{\frac{5\sigma}{2}}\rho_{x}\left(  \int_{x-\sigma}^{\frac{3\sigma}{2}}%
\frac{\rho_{y}+\rho_{y+\sigma}}{\lambda_{y}}dy\right)  dx\nonumber\\
&  -\sum_{a=0}^{2}\int_{\frac{\sigma}{2}+a\sigma}^{\frac{\sigma}{2}%
+a\sigma+\Delta}\rho_{x}\left(  \beta\phi_{x}+\ln\rho_{x}\right)  dx\nonumber
\end{align}
We again focus on the third line which we call $J$,%

\begin{align}
J  &  =\int_{\frac{\sigma}{2}+\Delta}^{\frac{3\sigma}{2}}\rho_{x}\left(
\int_{\frac{\sigma}{2}+\Delta}^{x}\frac{\rho_{y+\sigma}+\rho_{y}}{\lambda_{y}%
}dy\right)  dx+\int_{\frac{3\sigma}{2}+\Delta}^{\frac{5\sigma}{2}}\rho
_{x}\left(  \int_{x-\sigma}^{\frac{3\sigma}{2}}\frac{\rho_{y}+\rho_{y+\sigma}%
}{\lambda_{y}}dy\right)  dx\\
&  =\int_{\frac{\sigma}{2}+\Delta}^{\frac{3\sigma}{2}}\left(  \frac{d}{dx}%
\int_{\frac{\sigma}{2}+\Delta}^{x}\rho_{y}dy\right)  \left(  \int
_{\frac{\sigma}{2}+\Delta}^{x}\frac{\rho_{y+\sigma}+\rho_{y}}{\lambda_{y}%
}dy\right)  dx-\int_{\frac{3\sigma}{2}+\Delta}^{\frac{5\sigma}{2}}\left(
\frac{d}{dx}\int_{x}^{\frac{5\sigma}{2}}\rho_{y}dy\right)  \left(
\int_{x-\sigma}^{\frac{3\sigma}{2}}\frac{\rho_{y}+\rho_{y+\sigma}}{\lambda
_{y}}dy\right)  dx\nonumber\\
&  =\left(  \int_{\frac{\sigma}{2}+\Delta}^{\frac{3\sigma}{2}}\rho
_{y}dy\right)  \left(  \int_{\frac{\sigma}{2}+\Delta}^{\frac{3\sigma}{2}}%
\frac{\rho_{y+\sigma}+\rho_{y}}{\lambda_{y}}dy\right)  -\int_{\frac{\sigma}%
{2}+\Delta}^{\frac{3\sigma}{2}}\left(  \int_{\frac{\sigma}{2}+\Delta}^{x}%
\rho_{y}dy\right)  \frac{\rho_{x+\sigma}+\rho_{x}}{\lambda_{x}}dx\nonumber\\
&  +\left(  \int_{\frac{3\sigma}{2}+\Delta}^{\frac{5\sigma}{2}}\rho
_{y}dy\right)  \left(  \int_{\frac{\sigma}{2}+\Delta}^{\frac{3\sigma}{2}}%
\frac{\rho_{y}+\rho_{y+\sigma}}{\lambda_{y}}dy\right)  -\int_{\frac{3\sigma
}{2}+\Delta}^{\frac{5\sigma}{2}}\left(  \int_{x}^{\frac{5\sigma}{2}}\rho
_{y}dy\right)  \frac{\rho_{x-\sigma}+\rho_{x}}{\lambda_{x-\sigma}}dx\nonumber
\end{align}
or%
\begin{align}
J  &  =\left(  \int_{\frac{\sigma}{2}+\Delta}^{\frac{3\sigma}{2}}\rho
_{y}dy+\int_{\frac{3\sigma}{2}+\Delta}^{\frac{5\sigma}{2}}\rho_{y}dy\right)
\left(  \int_{\frac{\sigma}{2}+\Delta}^{\frac{3\sigma}{2}}\frac{\rho
_{y+\sigma}+\rho_{y}}{\lambda_{y}}dy\right) \\
&  -\int_{\frac{\sigma}{2}+\Delta}^{\frac{3\sigma}{2}}\left(  \int
_{\frac{\sigma}{2}+\Delta}^{x}\rho_{y}dy\right)  \frac{\rho_{x+\sigma}%
+\rho_{x}}{\lambda_{x}}dx-\int_{\frac{3\sigma}{2}+\Delta}^{\frac{5\sigma}{2}%
}\left(  \int_{x}^{\frac{5\sigma}{2}}\rho_{y}dy\right)  \frac{\rho_{x-\sigma
}+\rho_{x}}{\lambda_{x-\sigma}}dx\nonumber
\end{align}
This idea is to combine the last two terms to get things that look like
$\lambda_{x}$,%
\begin{align}
&  -\int_{\frac{\sigma}{2}+\Delta}^{\frac{3\sigma}{2}}\left(  \int
_{\frac{\sigma}{2}+\Delta}^{x}\rho_{y}dy\right)  \frac{\rho_{x+\sigma}%
+\rho_{x}}{\lambda_{x}}dx-\int_{\frac{3\sigma}{2}+\Delta}^{\frac{5\sigma}{2}%
}\left(  \int_{x}^{\frac{5\sigma}{2}}\rho_{y}dy\right)  \frac{\rho_{x-\sigma
}+\rho_{x}}{\lambda_{x-\sigma}}dx\\
&  =-\int_{\frac{\sigma}{2}+\Delta}^{\frac{3\sigma}{2}}\left(  \int
_{\frac{\sigma}{2}+\Delta}^{x}\rho_{y}dy\right)  \frac{\rho_{x+\sigma}%
}{\lambda_{x}}dx-\int_{\frac{\sigma}{2}+\Delta}^{\frac{3\sigma}{2}}\left(
\int_{\frac{\sigma}{2}+\Delta}^{x}\rho_{y}dy\right)  \frac{\rho_{x}}%
{\lambda_{x}}dx\nonumber\\
&  -\int_{\frac{\sigma}{2}+\Delta}^{\frac{3\sigma}{2}}\left(  \int_{x+\sigma
}^{\frac{5\sigma}{2}}\rho_{y}dy\right)  \frac{\rho_{x}}{\lambda_{x}}%
dx-\int_{\frac{\sigma}{2}+\Delta}^{\frac{3\sigma}{2}}\left(  \int_{x+\sigma
}^{\frac{5\sigma}{2}}\rho_{y}dy\right)  \frac{\rho_{x+\sigma}}{\lambda_{x}%
}dx\nonumber\\
&  =-\int_{\frac{\sigma}{2}+\Delta}^{\frac{3\sigma}{2}}\left(  \int
_{\frac{\sigma}{2}+\Delta}^{x}\rho_{y}dy+\int_{x+\sigma}^{\frac{5\sigma}{2}%
}\rho_{y}dy\right)  \frac{\rho_{x+\sigma}}{\lambda_{x}}dx-\int_{\frac{\sigma
}{2}+\Delta}^{\frac{3\sigma}{2}}\left(  \int_{\frac{\sigma}{2}+\Delta}^{x}%
\rho_{y}dy+\int_{x+\sigma}^{\frac{5\sigma}{2}}\rho_{y}dy\right)  \frac
{\rho_{x}}{\lambda_{x}}dx\nonumber\\
&  =-\int_{\frac{\sigma}{2}+\Delta}^{\frac{3\sigma}{2}}\left(  \int
_{\frac{\sigma}{2}+\Delta}^{\frac{5\sigma}{2}}\rho_{y}dy-\int_{x}^{x+\sigma
}\rho_{y}dy\right)  \frac{\rho_{x+\sigma}}{\lambda_{x}}dx-\int_{\frac{\sigma
}{2}+\Delta}^{\frac{3\sigma}{2}}\left(  \int_{\frac{\sigma}{2}+\Delta}%
^{\frac{5\sigma}{2}}\rho_{y}dy-\int_{x}^{x+\sigma}\rho_{y}dy\right)
\frac{\rho_{x}}{\lambda_{x}}dx\nonumber\\
&  =-\int_{\frac{\sigma}{2}+\Delta}^{\frac{3\sigma}{2}}\left(  \int
_{\frac{3\sigma}{2}+\Delta}^{\frac{5\sigma}{2}}\rho_{y}dy+\lambda_{x}%
-\lambda_{\frac{\sigma}{2}+\Delta}\right)  \frac{\rho_{x+\sigma}}{\lambda_{x}%
}dx-\int_{\frac{\sigma}{2}+\Delta}^{\frac{3\sigma}{2}}\left(  \int
_{\frac{3\sigma}{2}+\Delta}^{\frac{5\sigma}{2}}\rho_{y}dy+\lambda_{x}%
-\lambda_{\frac{\sigma}{2}+\Delta}\right)  \frac{\rho_{x}}{\lambda_{x}%
}dx\nonumber
\end{align}
or%
\begin{equation}
=-\left(  \int_{\frac{3\sigma}{2}+\Delta}^{\frac{5\sigma}{2}}\rho
_{y}dy-\lambda_{\frac{\sigma}{2}+\Delta}\right)  \int_{\frac{\sigma}{2}%
+\Delta}^{\frac{3\sigma}{2}}\frac{\rho_{x+\sigma}+\rho_{x}}{\lambda_{x}%
}dx-\int_{\frac{\sigma}{2}+\Delta}^{\frac{3\sigma}{2}}\rho_{x+\sigma}%
dx-\int_{\frac{\sigma}{2}+\Delta}^{\frac{3\sigma}{2}}\rho_{x}dx
\end{equation}
So%
\begin{equation}
J=\left(  \int_{\frac{\sigma}{2}+\Delta}^{\frac{3\sigma}{2}}\rho_{y}%
dy+\lambda_{\frac{\sigma}{2}+\Delta}\right)  \left(  \int_{\frac{\sigma}%
{2}+\Delta}^{\frac{3\sigma}{2}}\frac{\rho_{y+\sigma}+\rho_{y}}{\lambda_{y}%
}dy\right)  -\int_{\frac{\sigma}{2}+\Delta}^{\frac{3\sigma}{2}}\left(
\rho_{x+\sigma}+\rho_{x}\right)  dx
\end{equation}
and%
\begin{align}
\beta F\left[  \rho\right]   &  =\left(  1-\int_{\frac{\sigma}{2}+\Delta
}^{\frac{3\sigma}{2}}\rho_{x}dx\right)  \ln e^{\beta\phi_{\frac{\sigma}%
{2}+\Delta}}\rho_{_{\frac{\sigma}{2}+\Delta}}+\left(  1-\int_{\frac{3\sigma
}{2}+\Delta}^{\frac{5\sigma}{2}}\rho_{x}dx\right)  \ln e^{\beta\phi
_{\frac{5\sigma}{2}}}\rho_{\frac{5\sigma}{2}}-\ln\lambda_{\frac{\sigma}%
{2}+\Delta}-\int_{\frac{\sigma}{2}+\Delta}^{\frac{3\sigma}{2}}\frac{\rho_{y}%
}{\lambda_{y}}dy\\
&  +\int_{\frac{\sigma}{2}}^{\frac{5\sigma}{2}+\Delta}\rho_{x}\ln\rho
_{x}dx+\frac{1}{2}\int_{\frac{\sigma}{2}+\Delta}^{\frac{3\sigma}{2}}\rho
_{x}\ln\frac{\lambda_{\frac{\sigma}{2}+\Delta}}{\lambda_{x}}dx+\frac{1}{2}%
\int_{\frac{3\sigma}{2}+\Delta}^{\frac{5\sigma}{2}}\rho_{x}\ln\frac
{\lambda_{\frac{3\sigma}{2}}}{\lambda_{x-\sigma}}dx\nonumber\\
&  +\frac{1}{2}\left(  \int_{\frac{\sigma}{2}+\Delta}^{\frac{3\sigma}{2}}%
\rho_{y}dy+\lambda_{\frac{\sigma}{2}+\Delta}\right)  \left(  \int
_{\frac{\sigma}{2}+\Delta}^{\frac{3\sigma}{2}}\frac{\rho_{y+\sigma}+\rho_{y}%
}{\lambda_{y}}dy\right)  -\frac{1}{2}\int_{\frac{\sigma}{2}+\Delta}%
^{\frac{3\sigma}{2}}\left(  \rho_{x+\sigma}+\rho_{x}\right)  dx\nonumber\\
&  -\sum_{a=0}^{2}\int_{\frac{\sigma}{2}+a\sigma}^{\frac{\sigma}{2}%
+a\sigma+\Delta}\rho_{x}\left(  \beta\phi_{x}+\ln\rho_{x}\right)  dx\nonumber
\end{align}
Writing the last term of the first line as
\begin{align}
\int_{\frac{\sigma}{2}+\Delta}^{\frac{3\sigma}{2}}\frac{\rho_{y}}{\lambda_{y}%
}dy  &  =\frac{1}{2}\int_{\frac{\sigma}{2}+\Delta}^{\frac{3\sigma}{2}}%
\frac{\rho_{y}+\rho_{y+\sigma}}{\lambda_{y}}dy+\frac{1}{2}\int_{\frac{\sigma
}{2}+\Delta}^{\frac{3\sigma}{2}}\frac{\rho_{y}-\rho_{y+\sigma}}{\lambda_{y}%
}dy\\
&  =\frac{1}{2}\int_{\frac{\sigma}{2}+\Delta}^{\frac{3\sigma}{2}}\frac
{\rho_{y}+\rho_{y+\sigma}}{\lambda_{y}}dy+\frac{1}{2}\ln\frac{\lambda
_{\frac{3\sigma}{2}}}{\lambda_{\frac{\sigma}{2}+\Delta}}\nonumber
\end{align}
this becomes%
\begin{align}
\beta F\left[  \rho\right]   &  =\left(  1-\int_{\frac{\sigma}{2}+\Delta
}^{\frac{3\sigma}{2}}\rho_{x}dx\right)  \ln\frac{e^{\beta\phi_{\frac{\sigma
}{2}+\Delta}}\rho_{_{\frac{\sigma}{2}+\Delta}}}{\lambda_{\frac{\sigma}%
{2}+\Delta}^{1/2}}+\left(  1-\int_{\frac{3\sigma}{2}+\Delta}^{\frac{5\sigma
}{2}}\rho_{x}dx\right)  \ln\frac{e^{\beta\phi_{\frac{5\sigma}{2}}}\rho
_{\frac{5\sigma}{2}}}{\lambda_{\frac{3\sigma}{2}}^{1/2}}\\
&  +\int_{\frac{\sigma}{2}}^{\frac{5\sigma}{2}+\Delta}\rho_{x}\ln\rho
_{x}dx-\frac{1}{2}\int_{\frac{\sigma}{2}+\Delta}^{\frac{3\sigma}{2}}\rho
_{x}\ln\lambda_{x}dx-\frac{1}{2}\int_{\frac{3\sigma}{2}+\Delta}^{\frac
{5\sigma}{2}}\rho_{x}\ln\lambda_{x-\sigma}dx\nonumber\\
&  +\frac{1}{2}\left(  \int_{\frac{\sigma}{2}+\Delta}^{\frac{3\sigma}{2}}%
\rho_{y}dy+\lambda_{\frac{\sigma}{2}+\Delta}-1\right)  \left(  \int
_{\frac{\sigma}{2}+\Delta}^{\frac{3\sigma}{2}}\frac{\rho_{y+\sigma}+\rho_{y}%
}{\lambda_{y}}dy\right)  -\frac{1}{2}\int_{\frac{\sigma}{2}+\Delta}%
^{\frac{3\sigma}{2}}\left(  \rho_{x+\sigma}+\rho_{x}\right)  dx\nonumber\\
&  -\sum_{a=0}^{2}\int_{\frac{\sigma}{2}+a\sigma}^{\frac{\sigma}{2}%
+a\sigma+\Delta}\rho_{x}\left(  \beta\phi_{x}+\ln\rho_{x}\right)  dx\nonumber
\end{align}
we can also write this as
\begin{align}
\beta F\left[  \rho\right]   &  =\left(  1-\int_{\frac{\sigma}{2}+\Delta
}^{\frac{3\sigma}{2}}\rho_{x}dx\right)  \ln\frac{e^{\beta\phi_{\frac{\sigma
}{2}+\Delta}}\rho_{_{\frac{\sigma}{2}+\Delta}}}{\lambda_{\frac{\sigma}%
{2}+\Delta}^{1/2}}+\left(  1-\int_{\frac{3\sigma}{2}+\Delta}^{\frac{5\sigma
}{2}}\rho_{x}dx\right)  \ln\frac{e^{\beta\phi_{\frac{5\sigma}{2}}}\rho
_{\frac{5\sigma}{2}}}{\lambda_{\frac{3\sigma}{2}}^{1/2}}\\
&  +\int_{\frac{\sigma}{2}}^{\frac{5\sigma}{2}+\Delta}\rho_{x}\ln\rho
_{x}dx-\frac{1}{2}\int_{\sigma+\Delta}^{2\sigma}\left(  \rho_{x-\frac{\sigma
}{2}}+\rho_{x+\frac{\sigma}{2}}\right)  \ln\lambda_{x-\frac{\sigma}{2}%
}dx\nonumber\\
&  +\frac{1}{2}\left(  \int_{\frac{\sigma}{2}+\Delta}^{\frac{3\sigma}{2}}%
\rho_{y}dy+\lambda_{\frac{\sigma}{2}+\Delta}-1\right)  \left(  \int
_{\frac{\sigma}{2}+\Delta}^{\frac{3\sigma}{2}}\frac{\rho_{y+\sigma}+\rho_{y}%
}{\lambda_{y}}dy\right)  -1\nonumber\\
&  -\sum_{a=0}^{2}\int_{\frac{\sigma}{2}+a\sigma}^{\frac{\sigma}{2}%
+a\sigma+\Delta}\rho_{x}\left(  \ln e^{\beta\phi_{x}}\rho_{x}-\frac{1}%
{2}\right)  dx\nonumber
\end{align}
and%
\begin{equation}
\lambda_{x-\frac{\sigma}{2}}=\lambda_{\frac{\sigma}{2}+\Delta}+\int
_{\frac{\sigma}{2}+\Delta}^{\frac{3\sigma}{2}+\Delta}\rho_{y}dy-\int
_{x-\frac{\sigma}{2}}^{x+\frac{\sigma}{2}}\rho_{y}dy.
\end{equation}
Let us assume that in the small $\frac{\Delta}{\sigma}$ limit, all quantities
are sufficiently well behaved. and recall that
\begin{equation}
\int_{\frac{\sigma}{2}}^{\frac{\sigma}{2}+\Delta}e^{-\beta\phi_{y}%
}dy=e^{-\beta\phi_{\frac{\sigma}{2}+\Delta}}\frac{1}{\rho_{_{\frac{\sigma}%
{2}+\Delta}}}\lambda_{\frac{\sigma}{2}+\Delta}%
\end{equation}
which implies that $\lambda_{\frac{\sigma}{2}+\Delta}=O\left(  \frac{\Delta
}{\sigma}\right)  $. Furthermore, since, on physical grounds,
\[
\lim_{\frac{\Delta}{\sigma}\rightarrow0}\int_{\frac{\sigma}{2}+\Delta}%
^{\frac{3\sigma}{2}}\rho_{x}dx=1
\]
one concludes that for nonzero $\frac{\Delta}{\sigma}$, the integral is equal
to $1+O\left(  \frac{\Delta}{\sigma}\right)  $. As a consequence, one can say
that
\begin{align}
\beta F\left[  \rho\right]   &  =\int_{\frac{\sigma}{2}}^{\frac{5\sigma}%
{2}+\Delta}\rho_{x}\ln\rho_{x}dx-\frac{1}{2}\int_{\sigma+\Delta}^{2\sigma
}\left(  \rho_{x-\frac{\sigma}{2}}+\rho_{x+\frac{\sigma}{2}}\right)
\ln\lambda_{x-\frac{\sigma}{2}}dx-1\nonumber\\
&  +O\left(  \frac{\Delta}{\sigma},\frac{\Delta}{\sigma}\ln\frac{\Delta
}{\sigma}\right) \nonumber
\end{align}

\bigskip

\end{document}